\documentclass[prb,twocolumn,showpacs,preprintnumbers,flushbottom]{revtex4-1}

\usepackage[utf8]{inputenc}
\usepackage{graphicx}
\usepackage{ifthen}
\usepackage{color}
\newcommand{\blue}{\color{blue}}
\newcommand{\magenta}{\color{magenta}}

\usepackage{latexsym}
\usepackage{amssymb}
\usepackage{amsmath}
\usepackage{bm}
\DeclareBoldMathCommand{\bfsigma}{\sigma}
\DeclareBoldMathCommand{\bftau}{\tau}
\usepackage{bbm}

\newcommand{\half}{{\textstyle\frac{1}{2}}}
\newcommand{\quarter}{{\textstyle\frac{1}{4}}}
\newcommand{\im}{\mathop{\rm Im}}
\newcommand{\re}{\mathop{\rm Re}}
\newcommand{\eexp}{{\rm e}^}
\newcommand{\tr}{\mathop{\rm Tr}}

\newcommand{\deltaTwo}{\Delta}
\newcommand{\rhoS}{\rho}
\newcommand{\eV}{V}

\newcommand{\beq}[1]{\begin{eqnarray}\ifthenelse{#1=-1}{\nonumber}
{\ifthenelse{#1=0}{}{\label{e#1}}}}
\newcommand{\eeq}{\end{eqnarray}}

\renewcommand{\baselinestretch}{1.05}
\begin{document}

\title{Spin entanglement via STM current}
\author{Baruch Horovitz}
\affiliation{Department of Physics, Ben Gurion University, Beer Sheva 84105, Israel}
\author{Carsten Henkel}
\affiliation{University of Potsdam, Institute of Physics and Astronomy, 14476 Potsdam, Germany}
\begin{abstract} We consider a system of two spins under a scanning tunneling microscope bias and derive its master equation. We find that the tunneling elements to the electronic contacts (tip and substrate) generate an exchange interaction between the spins, as well as a Dzyaloshinskii-Moriya interaction in the presence of spin-orbit coupling. The tunnel current spectrum then shows additional lines compared to conventional spin resonance experiments.
When the spins have degenerate Larmor frequencies and equal tunneling amplitudes (without spin-orbit), there is a dark state with vanishing decay rate. The coupling to the electronic environment generates significant spin-spin entanglement via the dark state, even if the initial state is non-entangled.
\end{abstract}
\maketitle

Intense efforts are currently devoted to the study of two qubits coupled to an environment, motivated by quantum information science. In particular it has been realized that dissipative dynamics due to qubits coupling to the same bath can be tuned to yield entangled states both in theory \cite{alharbi,kastoryano,martin-cano,ficek} and experiment \cite{barreiro,krauter}. In parallel there has been considerable effort in developing techniques of scanning tunneling microscopy (STM) to probe electron spin resonance (ESR) features. These ESR-STM studies are of two types: either monitoring the current power spectrum in a DC bias employing a non-magnetic tip\cite{manassen1,balatsky1,manassen2} (first type), or monitoring the DC current with a magnetic tip when an additional AC voltage is tuned to resonance conditions \cite{mulleger,baumann,willke}.

\begin{figure}[tbh]
\centerline{\includegraphics*[width=0.5\columnwidth]{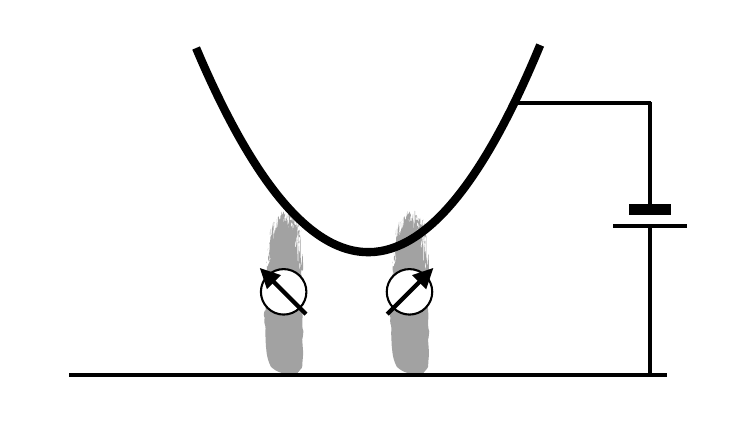}}
\caption[]{Sketch of the system: the spins may be thought as being located in two quantum dots or channels where electrons tunnel from a tip to the substrate, also denoted left (L) and right (R).}
\label{fig:sketch}
\end{figure}

In the present work we show that the STM setting with its two contacts provides {a novel} scenario for entangling two spins, representing two qubits. The presence of two non-degenerate spins has been proposed to account for the first type of ESR-STM phenomena.\cite{bh}
Here we study the case of degenerate spins, as sketched in Fig.\,\ref{fig:sketch}, which requires a new derivation of the appropriate master equation due to additional resonances. We find a number of  phenomena: (i) The tunneling couplings to the electronic baths (tip and substrate) generate dissipation, but also an exchange coupling between the two spins; in presence of spin-orbit coupling a  Dzyaloshinskii-Moriya interaction also emerges. (ii) The spin correlation functions as measured by an STM, in presence of either exchange or dipole-dipole interactions, show additional spectral lines relative to those in conventional ESR. (iii) When the tunneling amplitudes of the two spins are equal, we identify a dark state, i.e. an entangled state with infinite lifetime. An initial non-entangled state evolves into a significantly entangled state, i.e. environment-induced entanglement.

In the following we use a system+bath formalism where a system-environment interaction sums products of operators $A_j$, $B_j$ in the system and environment spaces, {respectively} \cite{shnirman1,schlosshauer}. We choose the $A_j$ such that they evolve in the interaction picture with frequency $\nu_j$:
\beq{01}
{\cal H}_{SE}(t) = \sum_{j} A_j B_j(t) \, \eexp{-i\nu_j t}
\eeq
where the sum may contain one or more terms with $\nu_j=0$. The master equation for the system density matrix $\rhoS$ is within the Born-Markov approximation
\beq{02}
\frac{d}{dt}\rhoS(t)
&=&\sum_{j,k} \Big\{
\tilde\Gamma_{jk}(\nu_k)\,\eexp{-i(\nu_j+\nu_k)t}
\nonumber
\\[-1ex]
&&
\phantom{\sum_{j,k} \Big\{}
{} \times [A_k\rhoS(t)A_j-A_j A_k\rhoS(t)]
+ \text{h.c.}
\Big\}
\\[-1ex]
\tilde\Gamma_{jk}(\omega)&=&\int_0^\infty d\tau \, \langle B_j(\tau)B_k(0)\rangle_E\,\eexp{i\omega \tau}
\nonumber
\eeq
where $\langle \ldots\rangle_E$ denotes the bath average. In the following we apply the secular approximation, i.e. only terms $j,k$ for which $\nu_j+\nu_k=0$ are kept, Eq.(\ref{e02}) then has the Lindblad form.~\cite{shnirman1,schlosshauer} This is justified when finite frequency differences are much larger than the linewidth. It is important to note that we do include off-diagonal terms in view of degeneracies in our system.

We {investigate} two isolated spins (quantum dots or impurities) described by Pauli matrices ${\bm\tau} \otimes \mathbbm{1}$ and $\mathbbm{1} \otimes \bm\tau$ (direct products display operators acting on the first spin times those acting on the second spin) {with} a common Larmor frequency $\nu$ {that are} coupled by tunneling in parallel to the two environments $L,\,R$ (Fig.\,\ref{fig:sketch}). The latter have spin-independent energies $\epsilon_{kL},\epsilon_{kR}$ whose chemical potentials differ by a bias $\eV$,
and Hamiltonian ${\cal H}_{L} = \sum_k \epsilon_{kL}c^\dagger_{kL}c_{kL}^{\phantom\dagger}$, with electron creation and annihilation operators $c^\dagger_{kL}, c_{kL}^{\phantom\dagger}$, being two-component spinors for each mode $kL$; similarly with $L \to R$.
Setting ${\cal H}_0 = \half\nu \, \tau_z\otimes \mathbbm{1} + \half\nu \, \mathbbm{1}\otimes\tau_z$  ($\hbar = 1$), the Hamiltonian is taken in the form
\beq{03}
{\cal H} &=& {\cal H}_0
+ {\cal H}_{L}
+ {\cal H}_{R}
\\
&& {} + \big(
  J_1 c_{R}^\dagger {\bm \sigma} c_{L}^{\phantom\dagger}
        \cdot {\bm\tau}\otimes \mathbbm{1}
+ J_2 c_{R}^\dagger {\bm \sigma} \hat u c_{L}^{\phantom\dagger}
        \cdot \mathbbm{1}\otimes{\bm\tau}
+ \text{h.c.}
  \big)
\nonumber
\eeq
where $c_L = \sum_k c_{kL}$ is the local operator that couples to the spins (same with $L \to R$).
The exchange tunneling terms in Eq.\,(\ref{e03}) are derived from tunneling via a localized state that has strong on-site Coulomb repulsion, which eliminates doubly occupied or zero occupied electron states, a procedure known as the Schrieffer-Wolff transformation.\cite{bh,hewson} For spin~2, we use the unitary matrix $\hat u = \eexp{i\sigma_z\phi}\eexp{i\sigma_y\theta/2}$ to model spin-orbit interactions; this is important for the coupling of an STM current to the spins.\cite{bh}
There are additional terms that tunnel electrons from one lead and back to the same lead, however the terms in~(\ref{e03}) dominate at large voltage, i.e. $\eV\gg \nu, k_BT$ ($T$ is temperature), the typical case in STM experiments.

The interaction picture relative to ${\cal H}_0 + {\cal H}_L + {\cal H}_R$
leads to ${\cal H}_{SE}(t)$ in the form~(\ref{e01}) with (using $\tau_\pm = \half(\tau_x \pm i\tau_y)$, $\epsilon_{RL} = \epsilon_{R} - \epsilon_{L}$ and an implicit summation over the bath levels $k$)
\begin{align}{\label{e04}}
A_1 &=\tau_-\otimes \mathbbm{1}\,, \qquad \nu_1 = \nu
\nonumber\\
B_1&=2J_1(c_{R}^\dagger\sigma_+c_{L}\eexp{i\epsilon_{RL}t}+
c_{L}^\dagger\sigma_+c_{R}\eexp{-i\epsilon_{RL}t})\nonumber\\
A_z &=\tau_z\otimes \mathbbm{1} \,, \qquad \nu_z = 0
\nonumber\\
B_z&=J_1c_{R}^\dagger\sigma_zc_{L}\eexp{i\epsilon_{RL}t}+ \text{h.c.}\nonumber\\
A_2 &=\mathbbm{1} \otimes \tau_- \,, \qquad\nu_{2} = \nu
\nonumber\\
B_2&=2J_2(c_{R}^\dagger\sigma_+\hat u c_{L}\eexp{i\epsilon_{RL}t}+
c_{L}^\dagger\hat u^\dagger\sigma_+ c_{R}\eexp{-i\epsilon_{RL}t})\nonumber\\
A_{z'} &=\mathbbm{1} \otimes \tau_z \,, \qquad \nu_{z'} = 0
\nonumber\\
B_{z'}&=J_2c_{R}^\dagger\sigma_z\hat u c_{L}\eexp{i\epsilon_{RL}t}+ \text{h.c.}
\end{align}
(Two additional terms are $A_{-j} = A_j^\dagger$ with $\nu_{-1,-2} = - \nu$.)
The product $A_z A_{z'}$ is secular and produces off-diagonal terms, as well as $A_1 A_{-2} = A_1 A_{2}^\dagger$ or $A_{-1} A_2 = A_{1}^\dagger A_2$. Plugging these expressions into Eq.\,(\ref{e02}) is straightforward and is detailed in the {Supplementary Material (SM)\cite{SM}}. Here we outline the form of one particular term
\beq{05}
&&
\frac{d\rhoS}{dt}=\ldots+\tilde\Gamma_{2,-1}(-\nu)
[A_{-1} \rhoS A_{2} - A_{2} A_{-1}\rhoS] + \text{h.c.}
\nonumber\\
&&\tilde\Gamma_{2,-1}(-\nu) = 4J_1J_2N^2(\epsilon_F)\cos\half\theta \, \eexp{-i\phi}
\\
&&
\phantom{\tilde\Gamma_{2,-1}(-\nu)=}
\times \int_{\epsilon_L, \epsilon_R} \hspace*{-1ex} i
\left\{\frac{f_R(1-f_L)}{\epsilon_{RL}-\nu+i0}-
\frac{f_L(1-f_R)}{\epsilon_{RL}+\nu-i0}\right\}
\nonumber
\eeq
where $\tr[\sigma_+\hat u\sigma_-]=\tr[\hat u^\dagger\sigma_+\sigma_-]=\cos\half\theta\,\eexp{-i\phi}$ is used, $f_R = f_R(\epsilon_R),f_L = f_L(\epsilon_L)$ are the Fermi distributions, containing the bias $\eV$, and $N^2(\epsilon_F)$ is the product of the two electronic densities of states at the Fermi energy.
We note that the principal part (P) of this integral is strongly cutoff-dependent, the cutoff $\Lambda$ being the electronic bandwidth on either tip or substrate, assumed comparable; the $\nu$ dependence in this term is weak provided $\nu\ll\Lambda$. The result is then, for $\eexp{-(\eV\pm\nu)/k_BT}\ll 1$:
\beq{06}
&& \tilde\Gamma_{2,-1}(-\nu) = 4 J_1J_2N^2(\epsilon_F)\cos\half\theta \, \eexp{-i\phi}
\nonumber\\
&& \phantom{\tilde\Gamma_{2,-1}(-\nu) = }
 \times [ \pi(\eV-\nu)-i\deltaTwo+i\nu\ln(\eV/\Lambda) ]
\\
&& \deltaTwo = {\rm P}\!\int_{\epsilon_R,\epsilon_L} \kern-1ex
\frac{f_L-f_R}{\epsilon_{R} - \epsilon_{L}}
\approx \Lambda\ln\frac{16\Lambda^2}{|\Lambda^2-V^2|}
-V\ln\left|\frac{\Lambda+V}{\Lambda-V}\right|
\nonumber
\eeq
This assumes constant densities of states, so that the expression for $\deltaTwo$ is taken just as an approximate indication that this term increases linearly with $\Lambda$ and therefore can be large.

Collecting all the secular terms that couple the two spins, i.e.\ $\tilde\Gamma_{jk}(\nu_k)$ with $(j,k)= (2,-1),\,(-2,1),\,(1,-2),\,(-1,2),\,(z,z'),\,(z',z)$ from the {SM \cite{SM}}, their imaginary parts combine into the effective interaction Hamiltonian
\beq{07}
{\cal H}_{\rm int} &=& -J_{\rm ex}\bm\tau_1\cdot\bm\tau_2+J_{\rm DM}[\bm\tau_1\times\bm\tau_2]_z
\nonumber\\
\left.
\begin{array}{r}
J_{\rm ex}\\
J_{\rm DM}
\end{array}
\right\}
&=& 4J_1J_2N^2(\epsilon_F)\deltaTwo\cos\half\theta
\left\{
\begin{array}{r}
\cos\phi\\
\sin\phi
\end{array}
\right.
\eeq
We recognize an exchange coupling as well as a Dzyaloshinskii-Moriya interaction. The latter appears in presence of spin-orbit coupling ($\phi\neq 0$) which breaks the symmetry between the two spins. We note that these bath-induced interactions are similar in spirit to the well known RKKY interaction that generates an exchange coupling between two separate spins in a metal, the metal being a common reservoir.\cite{kittel} Recall that the RKKY coupling is also sensitive to the cutoff as well as to the dimensionality of the metal.

\begin{figure*} \centering
\includegraphics*[width=.65\textwidth]{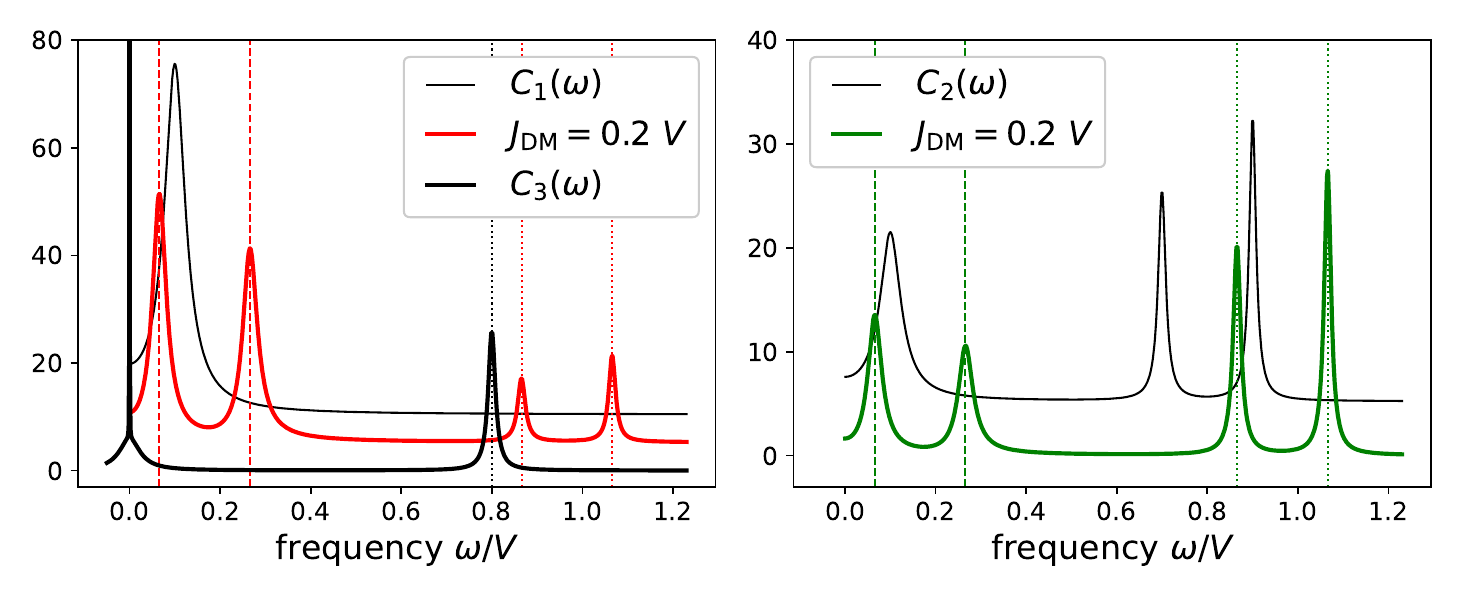}
\caption[]{
Correlations $C_1(\omega)$, $C_3(\omega)$ (\emph{left}) and $C_2(\omega)$ (\emph{right}) of Eq.\,(\ref{e08}) with the parameters:
Larmor {frequency} $\nu = 0.1\,\eV$ (including bath-induced shift), electrode couplings $\lambda_1=0.01,\,\lambda_2=0.009$ {where $\lambda_j=16\pi J_j^2N^2(\epsilon_F),\,(j=1,2)$,} spin-orbit angle $\theta = 0$ {and} exchange interaction $J_{\rm ex}=0.2\,\eV$.
Black lines: $\phi= 0$, no DM interaction; the peaks are at the expected positions
$\omega = \nu$ for $C_1(\omega)$, $\omega = \nu, 4J_{ex}\pm\nu$ for $C_2(\omega)$,
and $\omega = 0, 4 J_{\rm ex}$  for $C_3(\omega)$.
Thick colored lines: allowing for both exchange $J_{\rm ex} = 0.2\,\eV$ and DM interaction $J_{\rm DM} = 0.2\,\eV$ (i.e. $\phi = \pi/4$). The vertical dashed and dotted lines mark the transitions expected from the energy spectrum~(\ref{09}). $C_3(\omega)$ has a near $\delta(\omega)$ peak as well as a finite-width peak at $\omega=0$ whose widths correspond to the inverse lifetimes of the dark and bright states, respectively. Spectra are shifted vertically for clarity.}
\label{exchange}
\label{dipole_DM}
\end{figure*}

The environment-induced interaction can be detected in correlation functions that are measured by either ESR or an STM probe. Consider the correlations
\beq{08}
C_1(\omega) &=& \langle (\tau_-\otimes\tau_z+\tau_z\otimes\tau_-)_t(\tau_+\otimes\tau_z+\tau_z\otimes\tau_+)_0\rangle_\omega
\nonumber\\
&& \phantom{\langle (\tau_-\otimes\tau_z)_t(\tau_+\otimes\tau_z)_0\rangle_\omega}
+ (\omega\rightarrow -\omega)\nonumber
\\
C_2(\omega) &=& \langle (\tau_-\otimes\tau_z)_t(\tau_+\otimes\tau_z)_0\rangle_\omega +(\omega\rightarrow -\omega)
\nonumber\\
C_3(\omega) &=& \langle(\tau_-\otimes\tau_+)_t (\tau_+\otimes\tau_-)_0 \rangle_\omega + (\omega \rightarrow -\omega)
\eeq
Here $C_1(\omega)$ probes both spins equally, as in macroscopic ESR, while $C_2(\omega)$ probes only one spin, as allowed with STM. We recall that STM probes the spins via the current correlations \cite{balatsky1}, the current being $i J_1 c_{R}^\dagger {\bm \sigma} c_{L}^{\phantom\dagger}
        \cdot {\bm\tau}\otimes \mathbbm{1}
+ i J_2 c_{R}^\dagger {\bm \sigma} \hat u c_{L}^{\phantom\dagger}
        \cdot \mathbbm{1}\otimes{\bm\tau}
+ \text{h.c.}$
The spin-dependent current fluctuations therefore involve either $\tau_\pm$ for spin flip or $\tau_z$ otherwise. The current fluctuations allow also for a a double spin flip, hence detection of the correlation $C_3(\omega)$.
The correlation functions are computed from the quantum regression formula, {see SM \cite{SM}}.
To identify the various lines, we diagonalize the system Hamiltonian including ${\cal H}_{\rm int}$:
\begin{align}\label{09}
&
\frac{1}{\sqrt{2}}[\left|\uparrow\downarrow\right\rangle-\eexp{i\psi}\left|\downarrow\uparrow\right\rangle]:
&E_{S}& = 2\sqrt{J_{\rm ex}^2+J_{\rm DM}^2} + J_{\rm ex}
\nonumber\\
&\left|\uparrow\uparrow\right\rangle:
&E_{T1}&=\nu-J_{\rm ex}\nonumber\\
&
\frac{1}{\sqrt{2}}[\left|\uparrow\downarrow\right\rangle+\eexp{i\psi}\left|\downarrow\uparrow\right\rangle]:
&E_{T2}& = -2\sqrt{J_{\rm ex}^2+J_{\rm DM}^2} + J_{\rm ex}
\nonumber\\
&
\left|\downarrow\downarrow\right\rangle:
&E_{T3}&= -\nu-J_{\rm ex}
\end{align}
where $\tan\psi = J_{\rm DM} / J_{\rm ex}$. If both interactions are strictly within our model~(\ref{e07}), then remarkably, $\psi=\phi$, the spin-orbit phase. We use the labels S and T for singlet and triplet, although these coincide with the exact eigenstates only for $J_{\rm DM} = 0$.

Consider first the exchange-only case $\phi = 0$.
An ESR experiment allows only transitions within the triplet states, and all appear at frequency $\nu$. There are no transitions between the singlet and triplet states because of their opposite permutation symmetry, while the probing field is uniform in space {[see $C_1(\omega)$, Fig.\,\ref{exchange}(\emph{left})]}. In contrast, an STM experiment allows the probing current to tunnel via only one spin, hence permutation symmetry does not hold. The experiment would then show also singlet to triplet transitions, i.e. total of three lines at $\nu,\,\nu+4J_{\rm ex},\,|\nu-4J_{\rm ex}|$, as shown in Fig.\,\ref{exchange}(\emph{right}). Thus ESR-STM reveals the spectra of the two-spin system in more detail.

The case with spin-orbit coupling where both $J_{\rm ex}$ and $J_{\rm DM}$ are nonzero, is asymmetric within the pair (only the coupling to the second spin involves the spin-orbit matrix), hence both ESR and STM yield four lines: $\nu \pm 2 ( \sqrt{J_{\rm ex}^2 + J_{\rm DM}^2} - J_{\rm ex} )$ (the previous line at $\nu$ is split), and $2 ( \sqrt{J_{\rm ex}^2 + J_{\rm DM}^2} + J_{\rm ex} ) \pm \nu$.
This case is shown in Fig.\,\ref{dipole_DM}: $C_1(\omega)$ shows indeed four lines, although the additional two are rather weak. The STM case represented by $C_2(\omega)$ has four lines of comparable intensity.

The spectrum $C_3{(\omega)}$ shows {resonances at $\omega=0$ and} at $4J_{\rm ex}$ [Fig.\,\ref{dipole_DM}(\emph{left})]
because the operator $(\tau_- \otimes \tau_+) \rho_{\rm st}$ contains a superposition of S and T2 states with different energies.
A projection onto the entangled sub-space may be achieved by detecting {either of these resonances,} since $\tau_- \otimes \tau_+$ maps the non-entangled states $|{\uparrow\uparrow} \rangle$ and $|{\downarrow\downarrow} \rangle$ to zero.

\begin{figure*}[tbh]
\centering
\includegraphics*[height=.45\columnwidth]{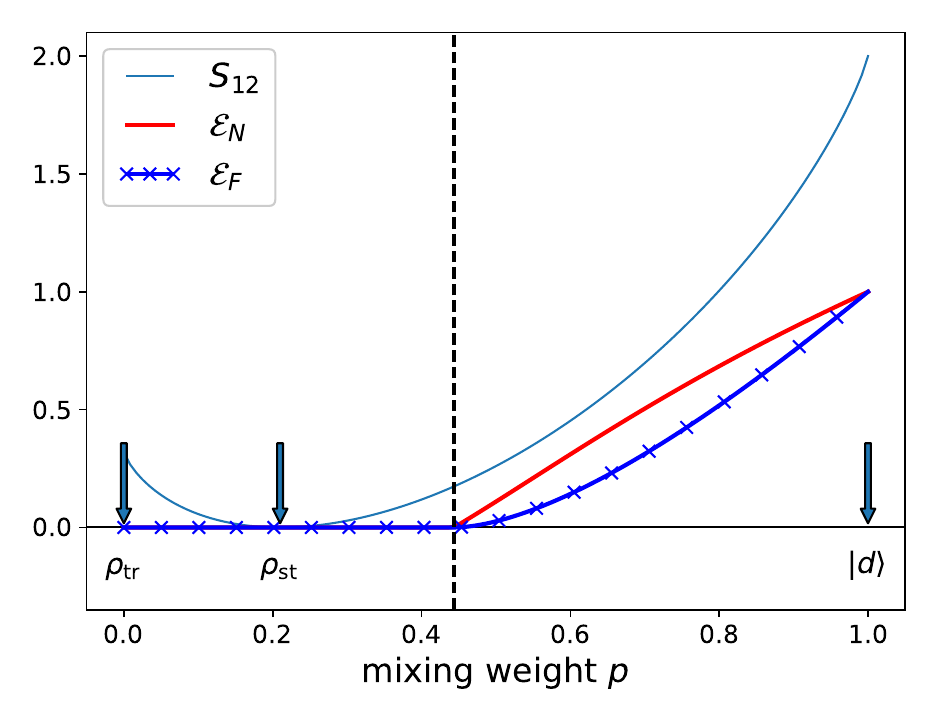}
\hspace*{05mm}
\includegraphics*[height=.45\columnwidth]{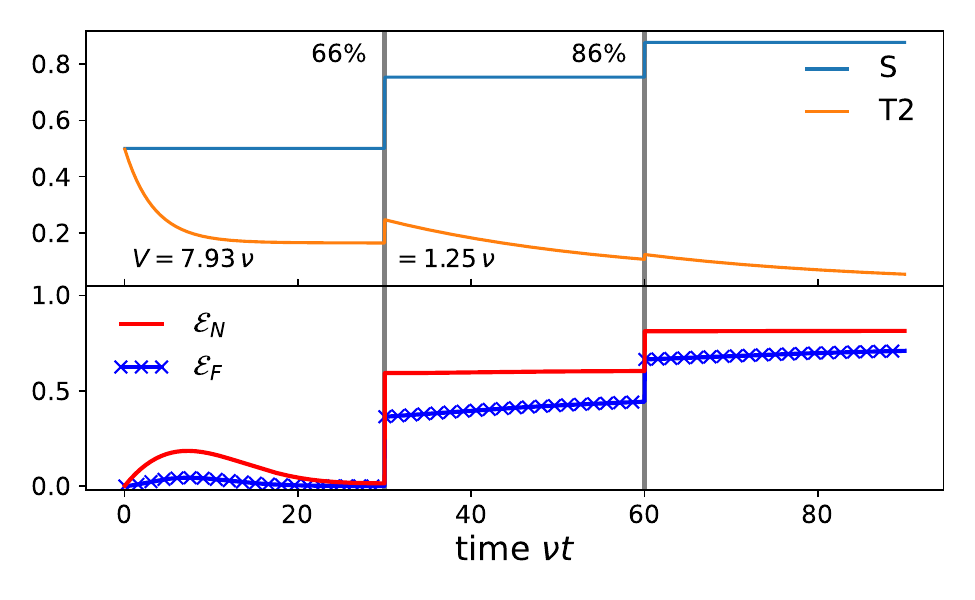}
\caption[]{
(\emph{left})
Entanglement measures at the symmetric point {$\lambda_1=\lambda_2,\, \theta = \phi=0$}: the steady state is given by a one-parameter family that interpolates between a mixed state $\rho_{\rm tr}$ in the triplet subspace (left, weight $1-p$) and the dark state (right, weight $p$). The third arrow marks the non-correlated stationary state $\rho_{\rm st}$ that is {also} found when the parameters are slightly detuning from the symmetric point.
Thin blue line: quantum mutual information $S_{12}$, nonzero nearly everywhere, but not specific to entanglement.
Thick red: logarithmic negativity ${\cal E}_N$,
thick blue with symbols: entanglement of formation ${\cal E}_F$,
dashed vertical line: critical mixing parameter $p_c$ (see main text).
Mutual information and entanglement measures are scaled to (e)bits, using logarithms to base $2$;
we take $\eV = 2.5\,\nu$.
(\emph{right})
Transient entanglement at the symmetric point, followed by probabilistic ``purification''.
Starting from the product state $|{\uparrow\downarrow}\rangle$, the entanglement of the
two-qubit system transiently goes through a maximum because the singlet state population (S) is constant, while the triplet sector relaxes to thermal equilibrium (only the S and T2 populations are shown; details about the transient entanglement such as oscillations and decoherence are discussed in the SM \cite{SM}).
The vertical lines mark a measurement of the $\tau_z \otimes \tau_z$ spin correlation.
Provided the measurement gives $-1$ (percentage given in top panel), the bias $\eV$ is decreased and the system relaxes to a lower effective temperature, increasing the entanglement.
A second measurement increases the entanglement further ($\eV$ is kept at the lower value).
The time evolution is computed from the eigenvalue spectrum of the master equation (see {SM \cite{SM}}).
Parameters: $\lambda = 0.0126$, $J_{\rm ex} \approx 0.024\,\nu$.
}
\label{dark}
\end{figure*}

We exhibit now the entanglement between the spins, induced by their interaction with the electronic bath, similar to works on two-qubit systems coupled by either a plasmonic waveguide \cite{martin-cano} or by cavity electrodynamics \cite{ficek}.
The most promising situation emerges when the spins are equally coupled to the environments, $J_1=J_2$, and $\theta = \phi = 0$ {(symmetric point)} so that spin-orbit does not distinguish between the spins. We then observe that the singlet state $| d \rangle = [\left|\uparrow\downarrow\right\rangle-\left|\downarrow\uparrow\right\rangle]/\sqrt{2}$ becomes `dark', meaning that it decouples from the three triplet states. For the dark state projector
\beq{10}
\hat d &=& \half
\big(\left|\uparrow\downarrow\right\rangle - \left|\downarrow\uparrow\right\rangle \big)
\big(\left\langle \uparrow\downarrow \right| - \left\langle\downarrow\uparrow\right| \big)
\\
&=& \quarter \mathbbm{1} \otimes  \mathbbm{1}
- \quarter \tau_z\otimes\tau_z
- \half\tau_-\otimes\tau_+
- \half\tau_+\otimes\tau_-
\nonumber
\eeq
we can show from the master equation that $d p_{d} / dt = \tr[\hat d (d\rho/dt)] = 0$ (see {SM \cite{SM}}), i.e., its decay rates vanish precisely. The stationary state {at this symmetric point} turns out to be a one-parameter family that interpolates between $\hat d$ and a mixture $\rho_{\rm tr}$ of triplet states with quasi-thermal populations (its effective temperature is $\approx \half \eV / k_B$ provided $\eV \gg \nu$). This remarkable phenomenon of a range of steady states is exhibited by
\beq{11}
\langle \tau_z \otimes \mathbbm{1} \rangle =
\langle \mathbbm{1} \otimes \tau_z \rangle &=&
- \tilde\nu(1 + 2\rho_{+-})
\\
\langle \tau_z \otimes \tau_z \rangle &=&
\tilde\nu^2 + 2(1+\tilde\nu^2)\rho_{+-}
\nonumber
\eeq
where $\tilde\nu=\nu/V$ and $\rho_{+-} = \langle \tau_+ \otimes \tau_- \rangle$ is an arbitrary {real} parameter, constrained only by the eigenvalues of $\rho$ being {between $0$ and $1$.}
The parameter $\rho_{+-}$ can also be detected from the $\delta(\omega)$ peak in the correlation
spectrum $C_3(\omega)$ [Eq.\,(\ref{e08}), in Fig.\,\ref{exchange}(\emph{left}) the near degenerate case is shown].
Fig.\,\ref{dark}(\emph{left}) quantifies the entanglement in this family of stationary states, as measured by the entanglement of formation ${\cal E}_F$ (related to the concurrence~\cite{Wootters_1998}) and the logarithmic negativity~\cite{Peres_1996,Horodecki_1996} ${\cal E}_N$. When the weight $p$ of the dark state exceeds $p_c = \half (1 - \tilde\nu^2)/(1 - \tilde\nu^2/3)$ (dashed line), there is stationary entanglement.

In Fig.\,\ref{dark}(\emph{right}), we show that an initially non-entangled state $|{\uparrow\downarrow}\rangle$ builds up entanglement during the relaxation of the triplet sector to equilibrium. We note that lowering $\eV$ towards $\nu$ (while keeping $\eexp{-(V-\nu)/k_BT}\ll 1$) decreases $p_c$:
a larger range of equilibrium states is then entangled.
The weight of the dark state does not exceed $p = \tfrac12$ with this easy-to-prepare product state, this is why we propose in Fig.\,\ref{dark}(\emph{right})
a route of further increasing the entanglement.
At the vertical gray lines, a measurement of the spin-spin correlation $\tau_z\otimes\tau_z$ is made:
from the result $-1$, the experimenter may infer that the two spins are not in the product states $|{\uparrow\uparrow}\rangle$ or $|{\downarrow\downarrow}\rangle$, while not destroying the relative phase of any entangled state.
A successful result thus increases the relative weight of the dark state and the entanglement.
Given the result $-1$ (antiparallel spins), the bias voltage is reduced, and the system relaxes to a lower effective temperature (in fact all relaxation rates decrease as well), and the entanglement increases. The measurement can be repeated and, if successful, purifies the dark state further.
To actually perform such a measurement,
one may monitor the $\omega=0$ peak of the current fluctuation spectrum $C_3(\omega)$. This peak signals the double spin-flip process, hence once observed the system collapses into the S or T2 states, equivalent to detecting an eigenvalue $-1$ of $\tau_z\otimes\tau_z$. We note that this procedure does not detect every double spin-flip since it should be made fast compared with the lifetime of the bright state so as to avoid generating the product states T1, T3.

In conclusion, we have shown a number of resonance phenomena that can be achieved by probing a pair of {degenerate} impurity spins in the tunnelling junction of an STM. These phenomena include generating exchange and DM interactions between the spins, the observation of additional lines in the STM setup, providing more information on the two-spin state. In some cases, a maximally entangled dark state emerges that is highly significant for quantum information applications.

\renewcommand{\baselinestretch}{1.1}

\begin{widetext}

\clearpage

\renewcommand{\beq}[1]{\begin{eqnarray}\ifthenelse{#1=-1}{\nonumber}
{\ifthenelse{#1=0}{}{\label{supp:e#1}}}}

\begin{center}
\textbf{\large Spin entanglement via STM current \\ \vspace{1mm}
Supplementary Material
\vspace{08mm}}
\\
\textrm{Baruch Horovitz{$^1$} and Carsten Henkel{$^2$} }
\\
\textit{{$^1$} Department of Physics, Ben
Gurion University of the Negev, Beer Sheva 84105, Israel}
\\
\textit{{$^2$}Institut f\"{u}r Physik, Karl-Liebknecht-Str.\ 24-25,
Universit\"{a}t Potsdam, 14476 Potsdam,
Germany}
\end{center}

We present in this supplementary a derivation of the master equation for various spin configurations, a brief discussion of the influence of a dipole-dipole interaction on the electron spin resonance, and details needed to evaluate correlation functions.

\subsection{Single spin}\label{s:single-spin}

As a preliminary, consider the Hamiltonian for a single spin coupled to electron baths in the left and right contacts, see also Ref.\,\onlinecite{sm:bh}.
The strong Coulomb interaction on the spin site allows virtual tunneling that involves the spins of the tunneling electrons (Pauli operator $\bfsigma$) and the local spin (Pauli operator $\bftau$), derived via the {Schrieffer}-Wolff transformation~\cite{sm:hewson}
\beq{14}
{\cal H}=\half\nu\tau_z
+ {\cal H}_{L} + {\cal H}_{R}
+ \big( J c_L^\dagger \bfsigma c_R^{\phantom\dagger}\cdot\bftau + \text{h.c.} \big)
\eeq
Here $\nu$ is the Larmor frequency ($\hbar = 1$), and
${\cal H}_L=\sum_k\epsilon_{kL}c^\dagger_{kL}c_{kL}^{\phantom\dagger},\, {\cal H}_R=\sum_k\epsilon_{kR}c^\dagger_{kR}c_{kR}^{\phantom\dagger}$ describe the environment (left and right electronic contacts). We use spinors $c^\dagger, c$ (between which $\bfsigma$ is acting); the energy levels $\epsilon_{kL},\epsilon_{kR}$ differ by a potential $\eV$. The tunneling is dominated by a single site so that $c_L=\sum_k c_{kL},\,c_R=\sum_k c_{kR}$. In general there are additional exchange terms of the form $c_R^\dagger\bfsigma c_R^{\phantom\dagger}\cdot \bftau,\, c_L^\dagger\bfsigma c_L^{\phantom\dagger}\cdot\bftau$, however these lead to relaxation rates $\sim k_B T$ (the temperature) while the term in \eqref{supp:e14} leads to $\sim \eV$, hence for the interesting case $\eV\gg k_B T$, the latter dominates.

The interaction picture is generated by the evolution operator
\beq{15}
U_e &=& \exp{\big[{-}i(\half\nu\tau_z + {\cal H}_L + {\cal H}_R)t \big]}
\eeq
which generates expressions like
$U_e^\dagger \tau_- U_e=\tau_- \, \eexp{-i\nu t}$
and
$U_e^\dagger c_L^\dagger c_R^{\phantom\dagger} U_e = c_L^\dagger c_R^{\phantom\dagger} \,\eexp{-i\epsilon_{RL}t}$
with $\epsilon_{RL}=\epsilon_R-\epsilon_L$ and an implicit summation over $k$ in the bath operators.
Using $\bfsigma \cdot\bftau =2\sigma_+\tau_-+2\sigma_-\tau_++\sigma_z\tau_z$ the system-environment (SE) coupling [second term of Eq.\,(\ref{supp:e14})] in the interaction picture becomes
\beq{16}
{\cal H}_{SE} =
2 J c_L^\dagger \sigma_+ c_R\tau_- \, \eexp{-i\epsilon_{RL}t-i\nu t}
+
2 J c_L^\dagger \sigma_- c_R\tau_+ \, \eexp{-i\epsilon_{RL}t+i\nu t}
+
  J c_L^\dagger \sigma_z c_R\tau_z \, \eexp{-i\epsilon_{RL}t}
+ \text{h.c.}
\eeq
We define the operators
\begin{align}{\label{supp:e17}}
A_1   &= \tau_- &
\nu_1 &= \nu &
B_1   &= 2 J (
c_L^\dagger\sigma_+c_R^{\phantom\dagger} \, \eexp{-i\epsilon_{RL}t}
+
c_R^\dagger\sigma_+c_L^{\phantom\dagger} \, \eexp{i\epsilon_{RL}t}
)
\nonumber
\\
A_z   &= \tau_z &
\nu_z &= 0 &
B_z   &=
J c_L^\dagger\sigma_z c_R^{\phantom\dagger} \, \eexp{-i\epsilon_{RL}t}
+ \text{h.c.}
\end{align}
with $A_{-1} = \tau_+$ and $B_{-1} = B_1^\dagger$ at $\nu_{-1} = -\nu$.
To second order in ${\cal H}_{SE}$,
the Bloch-Redfield master equation~\cite{sm:Bloch_1957, sm:Redfield_1957, sm:AlickiLendi_book} for the reduced density operator $\rho$ of the spin then takes the form [Eq.\,(\ref{e02}) in the main text]
\beq{0200}
\frac{d}{dt}\rho(t)
&=&\sum_{j,k}\tilde\Gamma_{jk}(\nu_k)\,\eexp{-i(\nu_j+\nu_k)t}[A_k\rho(t)A_j-A_j A_k\rho(t)]+ \text{h.c.}
\\
\tilde\Gamma_{jk}(\omega)&=&\int_0^\infty \!ds \, \langle B_j(s)B_k(0)\rangle_E\,\eexp{i\omega s}\nonumber
\eeq
where the correlation functions $\langle B_j(t')B_k(t)\rangle_E$ are taken with respect to the equilibrium state of the contacts and therefore depend only on the difference $s = t' - t$.

\subsubsection{Bath correlation spectra}

The correlation $\langle B_1(t') B_{-1}(t) \rangle_E = \Gamma_{1,-1}(t' - t)$
characterizes the `transverse' spin fluctuations of the electron bath (the ladder operators are linear combinations $\sigma_{\pm} = \half(\sigma_x \pm i \sigma_y)$).
Summing over the spinor indices of $c_L$ and $c_R$, we get
\beq{18}
\Gamma_{1,-1}(t' - t) &=& 4J^2
\big\langle
(c_L^\dagger\sigma_+ c_R^{\phantom\dagger} \, \eexp{-i\epsilon_{RL}t'}
+
c_R^\dagger\sigma_+ c_L^{\phantom\dagger} \, \eexp{i\epsilon_{RL}t'}
)
(c_R^\dagger\sigma_- c_L^{\phantom\dagger} \, \eexp{i\epsilon_{RL}t}
+
c_L^\dagger\sigma_- c_R^{\phantom\dagger} \, \eexp{-i\epsilon_{RL}t}
)
\big\rangle_E
\nonumber
\\
&=&
f_L(\epsilon_L)(1-f_R(\epsilon_R))\, \eexp{-i\epsilon_{RL}(t'-t)}
4J^2 \tr[\sigma_+\sigma_-]
+
f_R(\epsilon_R)(1-f_L(\epsilon_L))\, \eexp{i\epsilon_{RL}(t'-t)}
4J^2 \tr[\sigma_-\sigma_+]
\eeq
which has to be summed over the bath levels. Here $f_L(\epsilon_L) = \langle c_L^\dagger c_L^{\phantom\dagger} \rangle_E$ is the Fermi-Dirac distribution. The two traces are equal and yield
$\tr[\sigma_+\sigma_-] = 1$. The other relevant correlations are $\langle B_{-1}(t') B_{1}(t)\rangle_E$ and $\langle B_{z}(t') B_{z}(t) \rangle_E$. For all other combinations, one finds vanishing traces of Pauli matrices.

For the master equation~(\ref{supp:e0200}), we need the half-sided Fourier transform of $\Gamma_{1,-1}(t' - t)$.
Evaluating it with a convergence factor $\eexp{-\eta s}$ ($\eta \to 0$ through positive values),
and writing the summation over bath levels via the density of states $N(\epsilon)$,
one gets
\begin{equation}
\tilde\Gamma_{1,-1}(\omega) =
4 J^2N^2(0)\int\!{d\epsilon_R\,d\epsilon_L}\left\{
  \frac{f_L(\epsilon_L)(1-f_R(\epsilon_R))}{ \eta +i\epsilon_{RL}-i\omega }
+ \frac{f_R(\epsilon_R)(1-f_L(\epsilon_L))}{ \eta -i\epsilon_{RL}-i\omega }
\right\}
\label{supp:e19a}
\end{equation}
We assume that $N(\epsilon) = N(0)$ is constant, valid for a bandwidth much larger than the voltage $\eV$.

Consider first the real part that contains, in the limit $\eta \to 0$, $\delta$-functions. It will eventually determine transition rates, here due the transverse components of the bath spins:
\beq{19}
\re\tilde\Gamma_{1,-1}(\omega)
&=& 4
\pi J^2N^2(0)\int\!{d\epsilon_L}\Big\{f_L(\epsilon_L)(1-f_R(\epsilon_L+\omega))+
f_R(\epsilon_L-\omega)(1-f_L(\epsilon_L))\Big\}
\nonumber
\\
&=& \tfrac14 \lambda
\left\{
(\eV + \omega) \frac{ \eexp{\beta(\eV+\omega)} }{ \eexp{\beta(\eV+\omega)} - 1}
+
(\eV - \omega)\frac{ 1 }{ \eexp{\beta(\eV-\omega)} - 1}
\right\}
\eeq
with the dimensionless coupling constant $\lambda = 16\pi J^2N^2(0)$ and $1/\beta = k_B T$.
The left contact has its Fermi level shifted up by the bias voltage $\eV$.
The same integral needs to be evaluated for all correlations, only the value of the frequency $\omega = 0, \pm\nu$ changes.
A plot of the curly bracket in Eq.\,(\ref{supp:e19}) is shown in Fig.\ref{fig:transition-rate}(\emph{left}):
up to a factor, the same dependence on $\omega$ appears for all correlations $\tilde\Gamma_{jk}(\omega)$.

We focus for the following calculations on the low-temperature regime
$\eexp{-\beta(\eV\pm\omega)}\ll 1$.
This condition simplifies the expressions, though it is not essential. Furthermore, it justifies the Markov assumption since for $\omega\lesssim \eV$, $\tilde{\Gamma}_{jk}(\omega)$ is weakly $\omega$ dependent and $\Gamma_{jk}(t' - t)$ is short ranged.
We find
\beq{20}
\re\tilde{\Gamma}_{1,-1}(\omega) &=&
\quarter \lambda (\eV + \omega) + O(\eexp{-\beta(\eV\pm\omega)})
\nonumber
\eeq
The imaginary part of $\tilde\Gamma_{1,-1}(\omega)$ gives a principal value integral
\beq{21}
\im\tilde\Gamma_{1,-1}(\omega) &=& -4J^2N^2(0)
\,{\rm P}\!\!\int
\!{d\epsilon_R\,d\epsilon_L}
\frac{f_L(\epsilon_L)(1-f_R(\epsilon_R))(\epsilon_{RL}+\omega)
-f_R(\epsilon_R)(1-f_L(\epsilon_L))(\epsilon_{RL}-\omega)}
{\epsilon_{RL}^2-\omega^2}
\nonumber
\\
&=& -\frac{ \lambda }{ 4\pi }  \,{\rm P}\!\!\int\!{d\epsilon_R\,d\epsilon_L}
\left\{
\epsilon_{RL}\frac{
f_L(\epsilon_L)-f_R(\epsilon_R)
}{
\epsilon_{RL}^2-\omega^2
}
+
\omega\frac{
f_L(\epsilon_L)(1-f_R(\epsilon_R)) +f_R(\epsilon_R)(1-f_L(\epsilon_L))
}{
\epsilon_{RL}^2-\omega^2
}
\right\}
\nonumber
\\
&=& -\frac{ \lambda }{ 4\pi }
\left\{
\deltaTwo + \omega\ln\frac{\eV}{\Lambda}
\right\}
+ O(\eexp{-\beta(\eV\pm\omega)})
\eeq
where a cutoff $\Lambda$ is needed that we take $\Lambda \gg \nu \ge \omega$.
The second term of $\im\tilde\Gamma_{1,-1}(\omega)$ is smaller than the resonance linewidth (see below), yet we keep it, neglecting only $O(\eexp{-\beta(\eV\pm\omega)})$ terms. The first term $\deltaTwo$ of $\im\tilde\Gamma_{1,-1}(\omega)$  strongly depends on cutoffs, this is why if suffices to evaluate it in leading order, neglecting the (quadratic) $\omega$ dependence. This term  becomes
\beq{01}
\deltaTwo &\approx&
{\rm P}\!\!\int\!{d\epsilon_R\,d\epsilon_L}\frac{f_L(\epsilon_L)-f_R(\epsilon_R)}{\epsilon_{RL}}
\nonumber\\
&=&
\int\limits_{-\Lambda}^{\Lambda}d\epsilon_Rf_R(\epsilon_R)
\,{\rm P}\!\!\int\limits_{-\Lambda'}^{\Lambda'}\frac{d\epsilon_L}{\epsilon_L-\epsilon_R}
-\int\limits_{-\Lambda'}^{\Lambda'}d\epsilon_Lf_L(\epsilon_L)
\,{\rm P}\!\!\int\limits_{-\Lambda}^{\Lambda}\frac{d\epsilon_R}{\epsilon_L-\epsilon_R}
\nonumber
\\
&=&
2\Lambda'\ln\frac{\Lambda+\Lambda'}{\Lambda'}+
\Lambda\ln\frac{(\Lambda+\Lambda')^2}{|\Lambda^2-\eV^2|}
-\eV\ln\left|\frac{\Lambda+\eV}{\Lambda-\eV}\right|
\eeq
where $\Lambda,\Lambda'$ are the cutoffs (bandwidths) of the R, L electrodes, respectively, and temperature is neglected. When $\Lambda,\Lambda'\gg \eV$ then  $\deltaTwo=2\Lambda'\ln\frac{\Lambda+\Lambda'}{\Lambda'}+2\Lambda\ln\frac{\Lambda+\Lambda'}{\Lambda}$, i.e. it diverges logarithmically if one of the cutoffs is large and linearly when both are large.
In the case of two spins (see below) this term leads to an RKKY type exchange interaction between the spins.

The other relevant correlations turn out to be the transverse $\langle B_{-1}(t') B_{1}(t) \rangle_E$ and the `longitudinal' correlation $\langle B_z(t') B_z(t) \rangle_E$. For any other combination of indices, the traces of the Pauli matrices vanishes.
The two transverse correlations are identical, while for the longitudinal one, we need the trace $\tr \sigma_z^2 = 2$.
Re-writing the bath spin in Cartesian components $\sigma_x$, $\sigma_y$, $\sigma_z$, these numbers imply that its fluctuations are isotropic.
Defining $\delta = \nu\ln\frac{\eV}{\Lambda}$, we finally get the following set of weight factors
\beq{2301}
\begin{array}[c]{rcl}\displaystyle
\tilde{\Gamma}_{1,-1}(-\nu) &=&
	\displaystyle
	\frac{ \lambda }{ 4 } (\eV - \nu) - \frac{ i\lambda }{ 4\pi }
(
\deltaTwo - \delta
)
\,,
\\[2ex]
\tilde{\Gamma}_{-1,1}(\nu) &=&
	\displaystyle
	\frac{ \lambda }{ 4 } (\eV + \nu) - \frac{ i\lambda }{ 4\pi }
(
\deltaTwo + \delta
)
\,,
\end{array}
\qquad
\tilde{\Gamma}_{zz}(0) =
\frac{ \lambda }{ 8 } \eV
-
\frac{ i \lambda }{ 8 \pi } \deltaTwo
\,.
\eeq

\begin{figure}[tbh]
\centerline{
\includegraphics*[width=80mm]{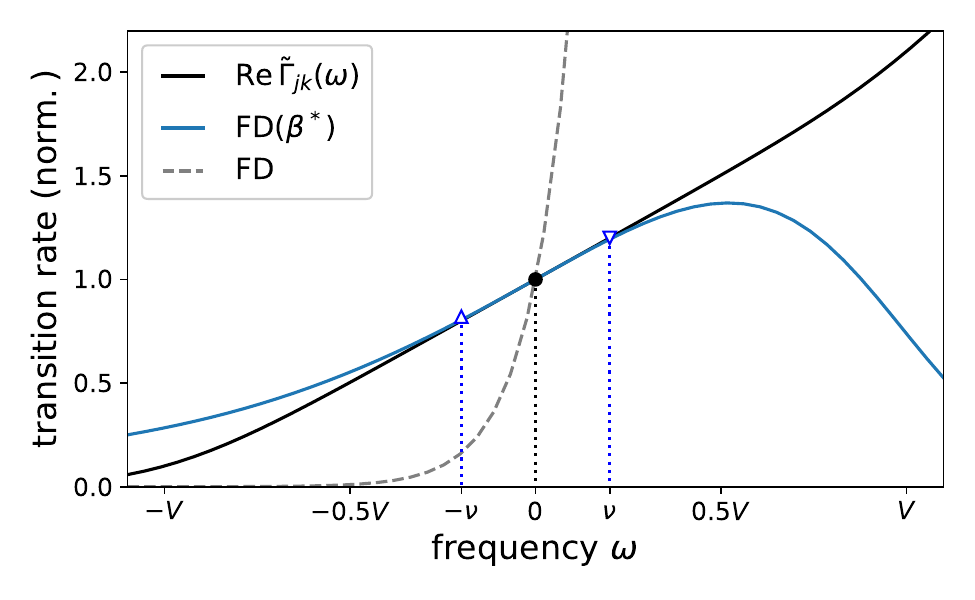}
\hspace*{03.5mm}
\includegraphics*[width=80mm]{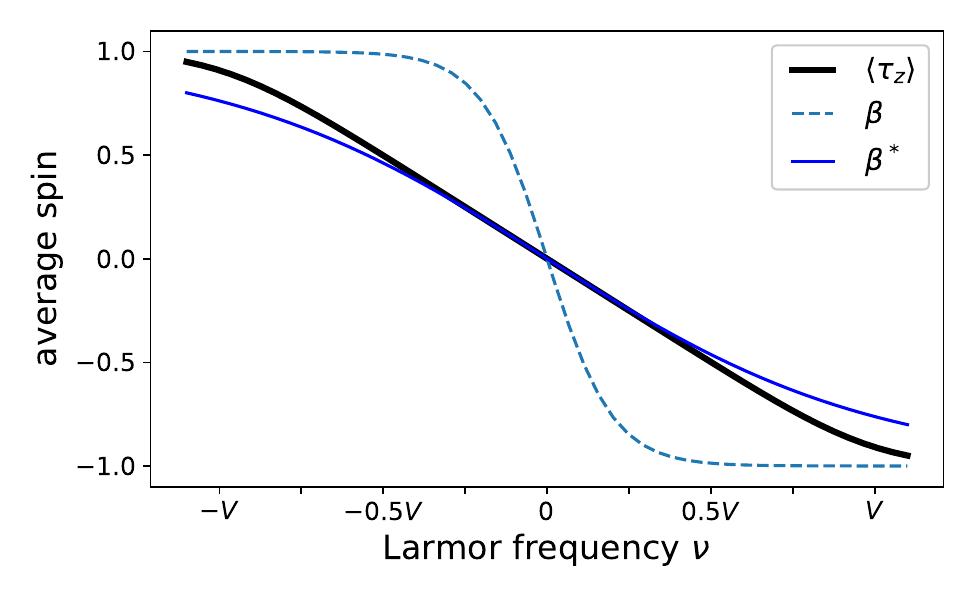}
}
\vspace*{-4ex}
\caption[]{\parbox[t]{140mm}{\raggedright
(\emph{left})
Spin relaxation rates as a function of the transition frequency.
We plot the expression in curly brackets in Eq.(\ref{supp:e19}), normalized
to its value for $\omega = 0$. Relevant values are
$\omega = 0$ (symbol $\bullet$, dephasing rate, contributes to transverse spin relaxation),
$\omega = \nu$ (symbols $\triangledown$, de-excitation rate, longitudinal spin relaxation),
$\omega = -\nu$ ($\triangle$, excitation).
The curve labelled FD$(\beta^*)$ represents
$\eexp{\beta^*\omega} \mathop{\rm Re} \tilde\Gamma_{jk}(-\omega)$, using a constant effective temperature $1/\beta^* = \half \eV$.
FD: same expression with the nominal temperature (we take $1/\beta = 0.1\,\eV$);
the large deviation illustrates that the fluctuation-dissipation relation does not hold for biased electronic contacts.
\\
(\emph{right}) Stationary value of the average impurity spin $\langle \tau_z \rangle$ (thick black line, Eq.\,(\ref{supp:e2502})), compared to thermal equilibrium predictions
$-\tanh(\half \nu / T)$ at the nominal temperature (curve $\beta$) and the effective temperature $\beta^*$. Same parameters as in the left panel.
}}
\label{fig:transition-rate}
\end{figure}

\subsubsection{Frequency shift and spin relaxation}

The master equation Eq.\,(\ref{supp:e0200}) becomes
\beq{23}
\frac{d\rho}{dt} &=& -\frac{i\nu}{2} [ \tau_z, \rho ]
\nonumber\\
&& {}
+
\left\{
  \tilde{\Gamma}_{1,-1}(-\nu)[\tau_+\rho\,\tau_- -\tau_-\tau_+\rho]
+ \tilde{\Gamma}_{-1,1}(\nu)[\tau_-\rho\,\tau_+ -\tau_+\tau_-\rho]
+ \tilde{\Gamma}_{zz}(0)[\tau_z\rho\,\tau_z -\tau_z^2\rho]
+
\text{h.c.}
\right\}
\eeq
We have written this in the laboratory picture, as can be seen from the commutator with the free spin Hamiltonian. This transformation removes from Eq.(\ref{supp:e0200}) the exponentials $\eexp{-i(\nu_j+\nu_k)t}$ (they arise from the free evolution of the operators $A_j A_k$).

Physical insight can be gained by considering first the imaginary parts of the coefficients $\tilde{\Gamma}_{jk}$.
Note that ``sandwich'' terms like $\tau_+\rho\,\tau_-$ and $\tau_z\rho\,\tau_z$ cancel when adding their h.c.
We then recognize a commutator
\beq{24}
&&
- \frac{ i\lambda }{ 4\pi }
\Big\{
(\deltaTwo-\delta)
\left[ -\tau_-\tau_+\rho
+\rho\,\tau_-\tau_+ \right]
+(\deltaTwo+\delta)
\left[ -\tau_+\tau_-\rho
+
\rho\,\tau_+\tau_- \right]
+ \half \deltaTwo
[ - \tau_z^2 \rho + \rho \tau_z^2 ]
\Big\}
\nonumber\\
&& \equiv -i[{\cal H}',\,\rho]
\qquad \text{with} \quad
{\cal H}'=-\frac{ \lambda }{ 4\pi } [(\deltaTwo-\delta)\tau_-\tau_++(\deltaTwo+\delta)\tau_+\tau_-
+ \half \deltaTwo \tau_z^2
]
=-\frac{ \lambda }{ 4\pi } [\delta\tau_z + \tfrac{3}{2}\deltaTwo \mathbbm{1}]
\eeq
Hence $\delta$ can be included into the Larmor frequency $\nu \mapsto \nu'$ (the shift being actually smaller than the linewidth), while $\tfrac{3}{2}\deltaTwo$ adds a mere constant to the effective Hamiltonian.

The real parts of $\tilde{\Gamma}_{jk}$ generate the dissipative terms in the master equation. They have the diagonal Lindblad structure
\begin{equation}
\frac{ d \rho }{ d t } =
- \frac{ i\nu' }{ 2 } \left[  \tau_z,\, \rho \right]
+
\sum_{l=\pm1,z} \gamma_l \left\{
L_l \rho L_l^\dag - \half L_l^\dag L_l \rho - \half \rho L_l^\dag L_l
\right\}
\label{supp:e2350}
\end{equation}
and we read off from Eq.\,(\ref{supp:e23}) the operators
\begin{align}{\label{supp:e2360}}
\hspace*{30mm}
L_{\pm1} &= \tau_\pm
	\,,
& L_z &= \tau_z
\hspace*{30mm}
\nonumber
\\
\gamma_{\pm1} &=
2 \re\tilde\Gamma_{\pm1,\mp1}(\mp\nu)
= \frac{ \lambda }{ 2 } (\eV \mp \nu)
\,,
& \gamma_{z} &= 2 \re\tilde\Gamma_{zz}(0)
= \frac{ \lambda }{ 4 } \eV
\end{align}
Evaluating the `sandwich' $L_{+1} \rho L_{+1}^\dagger$ with a pure state density operator, one recognizes that the rate $\gamma_{+1}$ describes the excitation $|{\downarrow} \rangle \to |{\uparrow} \rangle$ of the spin,
while $\gamma_{-1}$ describes relaxation down in energy. Getting back to the general expression~(\ref{supp:e19}), we get without bias
\begin{equation}
V = 0:\qquad
\gamma_{+1} =
\lambda \frac{\nu}{\eexp{\beta\nu}-1}
\,,\qquad
\gamma_{-1} =
\lambda \frac{\nu\,\eexp{\beta\nu}}{\eexp{\beta\nu} - 1}
\label{supp:e20a}
\end{equation}
The fluctuation-dissipation (FD) relation $\gamma_{+1} = \eexp{-\beta\nu}\gamma_{-1}$ holds with the bath temperature $1/\beta$, ensuring that the stationary spin populations correspond to thermal equilibrium set by $1/\beta$. For $V\neq 0$ and in the low-temperature regime $\eexp{-\beta(\eV\pm\nu)}\ll 1$ assumed in Eqs.\,(\ref{supp:e2360}), we rather have
\beq{20}
\gamma_{+1} &=& \eexp{-\beta^*\nu}\gamma_{-1}
\qquad \text{with} \qquad
\eexp{-\beta^*\nu}\equiv\frac{\eV-\nu}{\eV+\nu}
\eeq
Comparing now the ``up'' and ``down'' rates $\gamma_{+1}$, $\gamma_{-1}$,
we may interpret $1/\beta^*$ as an effective temperature for the spin populations,
although it depends in general on the frequency $\nu$.
If $\eV\gg \nu$, then $1/\beta^*\rightarrow \half\eV$.
In Fig.\,\ref{fig:transition-rate}(\emph{left}), we have plotted the up and down transition rates as a function of $\nu$ ($\omega = \mp\nu$ for up (down) transitions, respectively). The blue and dashed lines permit to check the FD relation: it does not hold at the nominal bath temperature (dashed) because the bias voltage brings the system out of equilibrium, and holds approximately if the effective $\beta^*$ is taken as a constant (blue).

While the Lindblad operators $\tau_\pm$ implement transitions between the spin states $|{\uparrow}\rangle$, $|{\downarrow}\rangle$, the operator $L_z = \tau_z$ in Eq.\,(\ref{supp:e2360}) describes a dephasing process that `scrambles' the relative phase of superpositions of $|{\uparrow}\rangle$, $|{\downarrow}\rangle$. This contributes to the relaxation of the transverse spin components $\tau_x$, $\tau_y$ and determines the  linewidth of the spin resonance.

Note the hierarchy of scales

\bigskip

$\qquad$
\begin{tabular}{ll}
$\nu$ & spin frequency, Zeeman splitting
\\
$\lambda\,\deltaTwo$ & global energy shift, exchange coupling [Eq.\,\ref{supp:e03}]
\\
$\lambda\,\eV$ & spin relaxation rate, linewidth [Eq.\,\ref{supp:e2502}]
\\
$\lambda\,\delta \sim \lambda\,\nu$ & frequency shift
\end{tabular}

\subsubsection{Bloch equations}
\label{supp:sec-Bloch-equations}

We arrive at the Bloch equations by expanding the density operator in the form
$\rho(t) = \half\mathbbm{1}+\rho_z(t)\tau_z+\rho_{+}(t)\tau_++\rho_{-}(t)\tau_-$.
Inserting this into Eq.(\ref{supp:e2350}), working out products of Pauli matrices and comparing coefficients, we find
\beq{2501}
\frac{d\rho_z}{dt} &=& -(\gamma_1+\gamma_{-1})\rho_z
+\half(\gamma_1-\gamma_{-1})=-\frac{1}{T_1}(\rho_z-\rho_z^0)
\nonumber
\\
\frac{d\rho_{+}}{dt} &=& - i \nu' \rho_{+}
- \half(\gamma_1 + \gamma_{-1}) \rho_+
- 2\gamma_{z} \rho_{+}
= - i \nu' \rho_{+} - \frac{1}{T_2}\rho_{+}
\eeq
[$\rho_{-}$ remains the complex conjugate of $\rho_{+}$, and $\nu' = \nu - (\lambda/2\pi) \delta$ includes the frequency shift from Eq.\,(\ref{supp:e24}).]
The relaxation times $T_1$ ($T_2$) for the longitudinal (transverse) spin components are given explicitly by
\beq{2502}
\frac{1}{T_1}&=&
\gamma_{1} + \gamma_{-1} =
\lambda\, \eV
,\qquad
\frac{1}{T_2}=\frac{1}{2T_1}+2\gamma_{z} =
\lambda\, \eV
\eeq
The stationary state is characterized by
$\rho_z^0=-\half\tanh\half\beta^*\nu = - \nu / \eV$.
It determines the population difference $\langle \tau_z\rangle_0 = 2 \rho_z^0 < 0$ for $\nu>0$, as expected. From \eqref{supp:e19}, the full dependence on parameters reads
[see Fig.\,\ref{fig:transition-rate}(\emph{right})]
\begin{equation}
\rho_z^0=\frac{-\nu}{\displaystyle\frac{\eV+\nu}{\tanh\half\beta(\eV+\nu)}+\frac{\eV-\nu}
{\tanh\half\beta(\eV-\nu)}}
\label{supp:e2504}
\end{equation}
a result known from studies of the Kondo model.\cite{sm:parcolet}

\subsection{Two spins}
\label{supp:two-spins}

Consider now two spins with
Larmor frequencies $\nu_1, \nu_2$.
Both spins are coupled by tunneling in parallel to the two environments $L,\,R$.
Using tensor product notation for the two sets of Pauli operators,
the Hamiltonian has the form
\beq{28}
{\cal H}=\half\nu_{1}\tau_z\otimes \mathbbm{1} + \half\nu_{2} \mathbbm{1} \otimes\tau_z
+
[ J_1 c_R^\dagger\bfsigma c_L\cdot\bftau \otimes \mathbbm{1}
+ J_2 c_R^\dagger\bfsigma \hat u c_L\cdot \mathbbm{1} \otimes\bftau + \text{h.c.} ]
+ {\cal H}_{R,L}
\eeq
Due to spin-orbit coupling, the bath spin is not exactly the same at the site of the second spin. This is represented by the unitary matrix $\hat u$ in the tunneling term through spin~2, as in earlier work by one of us.\cite{sm:bh}
In the following, we focus on the degenerate case $\nu_1 = \nu_2 = \nu$. Spins with finite detuning have been considered in Ref.\,\onlinecite{sm:bh}.

\subsubsection{Electron bath cross-correlations}
\label{supp:cross-correlations}

The interaction picture yields a time-dependent interaction
\beq{29}
{\cal H}_{SE}(t)
&=&
  2J_1 c_R^\dagger \sigma_+c_L\tau_-\otimes \mathbbm{1}\,\eexp{i\epsilon_{RL}t-i\nu_{1}t}
+ 2J_1 c_R^\dagger \sigma_-c_L\tau_+\otimes \mathbbm{1} \, \eexp{i\epsilon_{RL}t+i\nu_{1}t}
+  J_1 c_R^\dagger \sigma_zc_L\tau_z\otimes \mathbbm{1} \, \eexp{i\epsilon_{RL}t}
\nonumber
\\
&&
{}
+ 2J_2 c_R^\dagger \sigma_+\hat uc_L\otimes\tau_-\eexp{i\epsilon_{RL}t-i\nu_{2}t}
+ 2J_2 c_R^\dagger \sigma_-\hat uc_L\otimes\tau_+\eexp{i\epsilon_{RL}t+i\nu_{2}t}
+  J_2 c_R^\dagger \sigma_z\hat uc_L\otimes\tau_z\eexp{i\epsilon_{RL}t}
+ \text{h.c.}
\nonumber
\eeq
Hence ${\cal H}_{SE}$ has the form of Eq.\,(1) in the main text with
\begin{align}{\label{supp:e30}}
A_1 &=\tau_-\otimes \mathbbm{1} &\nu_{1}&=\nu
& B_1 &=2J_1(c_R^\dagger\sigma_+c_L\eexp{i\epsilon_{RL}t}+
c_L^\dagger\sigma_+c_R\eexp{-i\epsilon_{RL}t})
\nonumber
\\
A_{-1} &=\tau_+\otimes \mathbbm{1} &\nu_{-1}&=-\nu
& B_{-1}&=B_1^\dagger
\nonumber
\\
A_z &=\tau_z\otimes \mathbbm{1} &\nu_z&=0
& B_z&=J_1c_R^\dagger\sigma_zc_L\eexp{i\epsilon_{RL}t}+ \text{h.c.}
\nonumber
\\
A_2 &=\mathbbm{1} \otimes\tau_- &\nu_{2}&=\nu
& B_2&=2J_2(
c_R^\dagger\sigma_+\hat u c_L\eexp{i\epsilon_{RL}t}
+
c_L^\dagger\hat u^\dagger\sigma_+ c_R\eexp{-i\epsilon_{RL}t})
\nonumber
\\
A_{-2} &=\mathbbm{1} \otimes\tau_+ &\nu_{-2}&=-\nu
& B_{-2}&=B_2^\dagger
\nonumber
\\
A_{z'} &=\mathbbm{1} \otimes\tau_z &\nu_{z'}&=0
& B_{z'}&=J_2
c_R^\dagger\sigma_z\hat u c_L\eexp{i\epsilon_{RL}t}+ \text{h.c.}
\end{align}
The evaluation of the correlations $\Gamma_{jk}(t'-t) = \langle B_j(t') B_k(t) \rangle_E$ proceeds in the same way as in the previous section. We give below a few details for $\Gamma_{zz'}(t'-t)$ and $\Gamma_{2,-1}(t'-t)$.

In the secular approximation, only terms with $\nu_j + \nu_k = 0$ are kept (blue entries in Table~\ref{t:36coefficients}), which includes the spin-spin coupling generated by the correlation $\langle B_z B_{z'}\rangle$. For degenerate spins, also a correlation like $\langle B_{1} B_{-2}\rangle$ is secular (magenta in the Table). Terms like $\langle B_{z} B_{2}\rangle$ whose frequencies sum to $\nu$, much larger than the relevant line width, are neglected (black color).
We recall that the exponentials $\eexp{- i (\nu_j + \nu_k) t}$ in the master equation can be `transformed away' by reverting to the laboratory frame. This simply adds a commutator with the free spin Hamiltonian ${\cal H}_0$ to the equations of motion.

\begin{table}[bth]
\begin{tabular}{l|llllll}
$j \backslash k$ & $z$ & $z'$ & $+1$ & $+2$ & $-1$ & $-2$
\\
\hline
$z$ & {\blue $2$} & {\blue $2 \cos(\phi) c$} & $ $ & $-2 \,e^{-i \phi} s$ & $ $ & $-2 \,e^{i \phi} s$
\\
$z'$ & {\blue $2 \cos(\phi) c$} & {\blue $2$} & $2 \,e^{-i \phi} s$ & $ $ & $2 \,e^{i \phi} s$ & $ $
\\
$-1$ & $ $ & $2 \,e^{i \phi} s$ & {\blue $4$} & {\magenta $4 \,e^{-i \phi} c$} & $ $ & $ $
\\
$-2$ & $-2 \,e^{i \phi} s$ & $ $ & {\magenta $4 \,e^{i \phi} c$} & {\blue $4$} & $ $ & $ $
\\
$+1$ & $ $ & $2 \,e^{-i \phi} s$ & $ $ & $ $ & {\blue $4$} & {\magenta $4 \,e^{i \phi} c$}
\\
$+2$ & $-2 \,e^{-i \phi} s$ & $ $ & $ $ & $ $ & {\magenta $4 \,e^{-i \phi} c$} & {\blue $4$}
\\
\hline
$\omega$ & $0$ & $0$ & $\nu_1$ & $\nu_2$ & $-\nu_1$ & $-\nu_2$
\\
\hline
\end{tabular}
\caption[]{\parbox[t]{140mm}{\raggedright
Coefficients in the bath correlation functions
$\Gamma_{jk}(t' - t) = \langle B_j(t') B_k(t) \rangle_E$
arising from traces of Pauli matrices like
$\tr[\sigma_z^2 \hat{u}^\dag]$ and $\tr[\sigma_z^2 \hat{u}]$
in $\Gamma_{zz'}(t'-t)$ [Eq.(\ref{supp:e3101})]. Both terms evaluate
to the same number for all index pairs $(j,k)$.
Abbreviations:
$c = \cos(\theta/2)$, $s = \sin(\theta/2)$, using the Euler angle parametrization~(\ref{supp:u-parameters}) for the spin-orbit matrix $\hat u$.
The colored entries
correspond to secular (blue or magenta) terms for degenerate spins
$\nu_1 = \nu_2$. The bottom line gives the frequencies
$\omega = \nu_k$ where the
half-sided Fourier transforms $\tilde{\Gamma}_{jk}(\omega)$ need to be
evaluated.
}}
\label{t:36coefficients}
\end{table}

Among the correlations pertaining to different spins, consider first the longitudinal one
\beq{3101}
\Gamma_{zz'}(t)&=& \langle B_z(t) B_{z'}(0) \rangle_E
\nonumber
\\
&=& J_1J_2\langle(c_R^\dagger\sigma_z c_L\eexp{i\epsilon_{RL}t}
+c_L^\dagger\sigma_zc_R\eexp{-i\epsilon_{RL}t})(c_R^\dagger\sigma_z\hat u c_L+c_L^\dagger\hat u^\dagger \sigma_z c_R)\rangle
\nonumber
\\
&=& J_1J_2 \big\{
  f_R(\epsilon_R)(1-f_L(\epsilon_L))\eexp{i\epsilon_{RL}t}\tr[\sigma_z^2\hat u^\dagger]
+ f_L(\epsilon_L)(1-f_R(\epsilon_R))\eexp{-i\epsilon_{RL}t}\tr[\sigma_z^2\hat u]
\big\}
\eeq
The traces (see Table~\ref{t:36coefficients}) evaluate both to
$2\cos\half\theta\cos\phi$
when the spin-orbit matrix $\hat{u}$ is parametrized according to
\begin{equation}
\hat u = \exp({i\sigma_z\phi})\exp({i\sigma_y\theta/2})
\label{supp:u-parameters}
\end{equation}
The summation over the left and right bath levels leads to the same frequency dependence as in Eq.(\ref{supp:e19a}) so that the half-sided Fourier transform becomes
\beq{3102}
\tilde\Gamma_{zz'}(\omega) &=& 2J_1J_2N^2(0)\cos\half\theta\cos\phi
\int\!{d\epsilon_R\,d\epsilon_L}
\left\{
\frac{f_L(\epsilon_L)(1-f_R(\epsilon_R))}{\eta -i(\omega-\epsilon_{RL})}
+
\frac{f_R(\epsilon_R)(1-f_L(\epsilon_L))}{\eta -i(\omega+\epsilon_{RL})}
\right\}
\\
\tilde\Gamma_{zz'}(0) &=&
\Big(
\frac{\lambda_{12}}{8}
\eV
-
\frac{i\lambda_{12}}{8\pi}
\deltaTwo
\Big)
\cos\half\theta\cos\phi
+ O(\eexp{-\beta \eV})
\nonumber
\eeq
with $\lambda_{12} = 16\pi J_1 J_2 N^2(0)$
and $\deltaTwo$ defined in Eq.(\ref{supp:e01}).
For the other coefficient of this type, we find $\tilde\Gamma_{z'z}(0) = \tilde\Gamma_{zz'}(0)$.

The other nonzero off-diagonal terms are
$\tilde\Gamma_{2,-1}(-\nu)$,
$\tilde\Gamma_{-2,1}(\nu)$,
$\tilde\Gamma_{1,-2}(-\nu)$, and
$\tilde\Gamma_{-1,2}(\nu)$.
The first one, for example, arises from the correlation
$\langle B_{2}(t)B_{-1}(0)\rangle_E$ and involves the traces $4 \tr[\sigma_+\hat u\sigma_-] = 4 \tr[\hat u^\dagger\sigma_+\sigma_-] = 4 \cos\half\theta \, \eexp{-i\phi}$ (see Table~\ref{t:36coefficients}).
The remaining expression varies in the same way with frequency as Eq.\,(\ref{supp:e3102}). Evaluating at $\omega = -\nu$, we get within our approximations
\beq{3103}
\tilde\Gamma_{2,-1}(-\nu)
&=&
\Big[
\frac{\lambda_{12}}{4} (\eV - \nu)
-
\frac{i \lambda_{12}}{4\pi}(\deltaTwo - \delta)
\Big]
\cos\half\theta \, \eexp{-i\phi}
+ O(\eexp{-\beta(\eV\pm \nu)})
\eeq
where $\delta = \nu \ln\frac{\eV}{\Lambda}$.
The other coefficients have up to some sign flips the same structure and we list them below.

It is expedient to separate the set of twelve complex coefficients $\tilde{\Gamma}_{jk}( \nu_k )$ into those that generate an effective Hamiltonian, on the one hand, and those for relaxation processes of Lindblad type, on the other. With our notation for the operators of the two-spin system, we are led to
\beq{3100}
\tilde{\Gamma}_{jk}( \nu_k ) &=& \half \gamma_{jk} + i h_{jk}
\qquad
\Bigg\{
\begin{array}[c]{rcl}\displaystyle
h_{jk} &=& -\tfrac{ i }{ 2 } \big[
\tilde{\Gamma}_{jk}( \nu_k ) - \tilde{\Gamma}_{-k,-j}^*( -\nu_j )
\big]
\\[1ex]
\gamma_{jk} &=& \tilde{\Gamma}_{jk}( \nu_k ) + \tilde{\Gamma}_{-k,-j}^*( -\nu_j )
\end{array}
\eeq
(Recall that an index $-j$ corresponds to the hermitean conjugate operator $A_{-j} = A_{j}^\dagger$ and that the `zero indices' $j = z, z'$ are invariant, since $A_{z}$ and $A_{z'}$ are hermitean.)
In the one-spin case considered before, this splitting just corresponds to the real and imaginary parts of the $\tilde{\Gamma}_{jk}( \nu_k )$, and this is still true for two spins and the secular terms at hand, i.e., when $\nu_j + \nu_k = 0$. Otherwise, both $\gamma_{jk}$ and $h_{jk}$ are complex.

This construction makes the contributions of off-diagonal coefficients to the two-spin dynamics more transparent.
The bath-induced Hamiltonian, for example, is given by the expression
\beq{3110}
{\cal H}' &=& \sum_{j,k} h_{jk} A_j A_k
\eeq
This is indeed hermitean, since
${\cal H}^{\prime\dagger}
= \sum_{j,k} h^*_{jk} A_{-k} A_{-j}
= \sum_{j,k} h^*_{-k,-j} A_{j} A_{k}
= \sum_{j,k} h_{jk} A_{j} A_{k}$,
the last step being in virtue of Eq.\,(\ref{supp:e3100}).
By the same token, one checks that from the coefficients $i h_{jk}$ in the master equation, the `sandwich terms' $i \sum_{jk} h_{jk} A_k \rho A_j + \text{h.c.}$ mutually cancel.

The following calculation is actually simplified by the block structure visible in Table~\ref{t:36coefficients} when only secular terms are kept (blue and magenta):
groups of four terms can be identified in the double sum.
This is the structure we use in the following tables for the coefficients of the effective Hamiltonian.

We find with this approximation
\begin{equation}\label{supp:e-hminus1-array}
\begin{pmatrix}
h_{-1,1} & h_{-1,2}
\\
h_{-2,1} & h_{-2,2}
\end{pmatrix}
=
- \frac{ \deltaTwo + \delta }{4\pi}
\begin{pmatrix}
\lambda_1 & \lambda_{12} \cos(\half\theta) \, \eexp{-i\phi}
\\
\lambda_{12} \cos(\half\theta) \, \eexp{i\phi} & \lambda_2
\end{pmatrix}
=
- \frac{ \deltaTwo + \delta }{4\pi}
\mathbbm{B}
\end{equation}
In the next group, the frequency shift $\delta$ appears with the opposite sign (and the matrix $\mathbbm{B}$ is transposed)
\begin{equation}\label{supp:e-hplus1-array}
\begin{pmatrix}
h_{1,-1} & h_{1,-2}
\\
h_{2,-1} & h_{2,-2}
\end{pmatrix}
=
- \frac{ \deltaTwo - \delta }{4\pi}
\mathbbm{B}^{\sf T}
\end{equation}
These blocks~(\ref{supp:e-hminus1-array}, \ref{supp:e-hplus1-array}) can actually be combined because, for example,
the terms $h_{-1,2} (\tau_+ \otimes \tau_-) + h_{2,-1} (\tau_+ \otimes \tau_-)$
involve the same operator product.
The last piece involves the longitudinal spin operators and forms a real symmetric array
\begin{equation}\label{supp:e-hzz-array}
\begin{pmatrix}
h_{zz} & h_{zz'}
\\
h_{z'z} & h_{z'z'}
\end{pmatrix}
=
- \frac{ \deltaTwo }{16\pi}
\big( \mathbbm{B} + \mathbbm{B}^{\sf T} \big)
\end{equation}

For the `real parts', the corresponding tables with coefficients $\gamma_{jk}$ read
\begin{equation}\label{supp:gamma-minus1-array}
\begin{pmatrix}
\gamma_{-1,1} & \gamma_{-1,2}
\\
\gamma_{-2,1} & \gamma_{-2,2}
\end{pmatrix}
=
\frac{ \eV + \nu }{2}
\mathbbm{B}
\end{equation}
We shall see that these coefficients determine `jump down' processes, with a Lindbladian that combines the single-spin flip operators $\tau_- \otimes \mathbbm{1}$ and $\mathbbm{1} \otimes \tau_-$ [see Eq.\,(\ref{supp:eq_def-dark-state-from-down-jump}) in Sec.\,\ref{s:two-spin-jumps}].
The following set generates `jump up' processes:
\begin{equation}\label{supp:gamma-plus1-array}
\begin{pmatrix}
\gamma_{1,-1} & \gamma_{1,-2}
\\
\gamma_{2,-1} & \gamma_{2,-2}
\end{pmatrix}
=
\frac{ \eV - \nu }{2}
\mathbbm{B}^{\sf T}
\end{equation}
The last set will contribute to dephasing:
\begin{equation}\label{supp:gamma-zz-array}
\begin{pmatrix}
\gamma_{zz} & \gamma_{zz'}
\\
\gamma_{z'z} & \gamma_{z'z'}
\end{pmatrix}
=
\frac{ \eV }{8}
\big( \mathbbm{B} + \mathbbm{B}^{\sf T} \big)
\end{equation}

In the following sections, we deal separately with the bath-induced Hamiltonian and the dissipative processes in the master equation.

\subsubsection{Spin-spin interaction}

By a calculation identical to Eq.(\ref{supp:e24}), second line, the free Hamiltonian plus the diagonal terms $h_{j,-j}$ in Eqs.\,(\ref{supp:e-hminus1-array}--\ref{supp:e-hzz-array}) give the single-spin terms
\begin{equation}
{\cal H}_0' =
\half \nu_1' \tau_z \otimes \mathbbm{1}
+ \half \nu_2' \mathbbm{1} \otimes \tau_z
\qquad
\text{with}\quad
\nu_j' = \nu - \frac{\lambda_j}{2\pi} \delta
\quad (j = 1, 2)
\label{supp:eq_single-spin-effective-Hamiltonian}
\end{equation}
In addition, a global shift $-(3 / 8\pi) (\lambda_1 + \lambda_2) \deltaTwo$ appears that is not relevant in the following.

The coupling to the bath induces an effective interaction ${\cal H}'$ between the spins that is the sum of three terms.
The one given by the off-diagonal elements $h_{zz'}$ and $h_{z'z}$ in Eq.\,(\ref{supp:e-hzz-array}) is the simplest:
\beq{03}
{\cal H}_1' = - J_{\rm ex} \tau_z\otimes\tau_z,\qquad
\text{with}\quad
J_{\rm ex} =
\frac{\lambda_{12}}{4\pi} \deltaTwo \cos(\half\theta)\, \cos\phi
\eeq
The contributions of Eqs.\,(\ref{supp:e-hminus1-array}, \ref{supp:e-hplus1-array}) combine into
(the frequency shift $\delta$ drops out with opposite signs)
\beq{0402}
{\cal H}_2' & = &
  (h_{-1,2} + h_{2,-1}) (\tau_+\otimes\tau_-)
+ (h_{-2,1} + h_{1,-2}) (\tau_-\otimes\tau_+)
\nonumber
\\
&=&
- \frac{ \lambda_{12} }{ 2\pi } \deltaTwo \cos(\half\theta)
\left\{ \eexp{-i\phi}\,\tau_+\otimes\tau_- + \text{h.c.} \right\}
\nonumber
\\
&=&
- J_{\rm ex}
[ \tau_x\otimes\tau_x + \tau_y\otimes\tau_y ]
+ \frac{ \lambda_{12} }{ 4\pi } \deltaTwo \cos(\half\theta) \sin\phi
[ \tau_x\otimes\tau_y - \tau_y\otimes\tau_x ]
\eeq
The total induced interaction for the two spins can be written in the suggestive form
\beq{05}
{\cal H}' &=& {\cal H}_1' + {\cal H}_2' =
- J_{\rm ex} \bftau_1 \cdot \bftau_2
+ J_{\rm DM} [\bftau_1 \times \bftau_2]_z
\qquad
\text{with}\quad
J_{\rm DM} =
\frac{ \lambda_{12} }{ 4 \pi }
\deltaTwo\cos(\half\theta)\sin\phi
\eeq
where $J_{\rm DM}$ is a Dzyaloshinskii-Moriya coupling.\cite{sm:miyahara}
This anisotropic exchange term may have a noticeable shift on the observed resonance frequencies. It is similar to the exchange coupling of RKKY type between two separate spins in a homogeneous metal, where the conduction electrons provide a common reservoir.\cite{sm:kittel}

The interaction splits into sub-blocks when written as a matrix in the eigenbasis of ${\cal H}_0'$. The upper and lower states just shift down by $J_{\rm ex}$ under ${\cal H}_1'$ and are not perturbed by ${\cal H}_2'$. Introducing the coupling $J' = (\lambda_{12} / 2\pi )\deltaTwo \cos\half\theta
= 2 \sqrt{ J_{\rm ex}^2 + J_{\rm DM}^2 }$, the total effective Hamiltonian ${\cal H}'$ is represented in the $( | {\uparrow\uparrow} \rangle,| {\uparrow\downarrow} \rangle, \, | {\downarrow\uparrow} \rangle, \, | {\downarrow\downarrow} \rangle)$ basis by the block matrix (void elements vanish):
\begin{equation}
({\cal H}') =
\begin{pmatrix}
\half( \nu_1' + \nu_2' ) - J_{\rm ex} & & & \\
& \half( \nu_1' - \nu_2' ) + J_{\rm ex} & - J' \, \eexp{-i\phi} &
\\
& - J' \, \eexp{i\phi} & \half( \nu_2' - \nu_1' ) + J_{\rm ex} &
\\
 & & & - \half( \nu_2' + \nu_1' ) - J_{\rm ex}
\end{pmatrix}
\label{supp:eq-effective-H}
\end{equation}
With the help of the mixing angle
\begin{equation}
\tan\chi = \frac{ J' }{ \half( \nu_1' - \nu_2' ) }
{}
= - \frac{ \lambda_{12} }{ \half( \lambda_1 - \lambda_2 ) }
    \frac{ \Delta }{ \delta } \cos(\half\theta)
\label{supp:eq_def-mixing-H}
\end{equation}
the eigenstates and energy levels in the two-dimensional (`intermediate') subspace
follow from a standard calculation:
\begin{eqnarray}
| d \rangle =
\cos(\chi/2)
\eexp{ -i \phi / 2 } | {\uparrow\downarrow} \rangle
-
\sin(\chi/2)
\eexp{ i \phi / 2 } | {\downarrow\uparrow} \rangle
: \qquad && E_{d} =
J_{\rm ex}
+
\sqrt{ J^{\prime 2} + \tfrac{1}{4}(\nu_1' - \nu_2')^2}
\nonumber
\\
| b \rangle =
\sin(\chi/2)
\eexp{ -i \phi / 2 } | {\uparrow\downarrow} \rangle
+
\cos(\chi/2)
\eexp{ i \phi / 2 } | {\downarrow\uparrow} \rangle
: \qquad && E_{b} =
J_{\rm ex}
-
\sqrt{ J^{\prime 2} + \tfrac{1}{4}(\nu_1' - \nu_2')^2}
\label{supp:eigenstates}
\end{eqnarray}
This generalizes the spectrum given in Eq.(\ref{e10}) of the main text to allow for non-equal renormalized Larmor frequencies, $\nu_1' = \nu_2'$.
The energy spectrum is shown in Fig.\,\ref{suppfig:d-b-levels}(\emph{left}) vs.\ the coupling constant $\lambda_1$ at fixed $\lambda_2$. (This situation mimicks a spin localised near the STM tip that is approaching the site of the other spin.)
At equal couplings, $\chi = \pi/2$, and the eigenstates $| d \rangle$, $| b \rangle$
become maximally entangled, with a relative phase set by the spin-orbit coupling angle $\phi$.
We call these `dark' and `bright' states, which will be justified in the following.

\begin{figure}[thb]
\centerline{
\includegraphics*[height = 45mm]{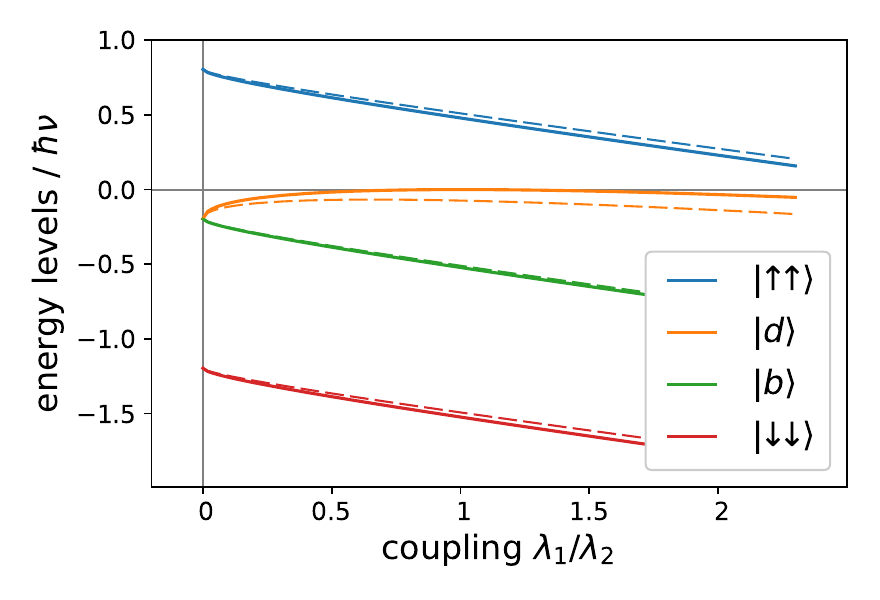}
\includegraphics*[height = 45mm]{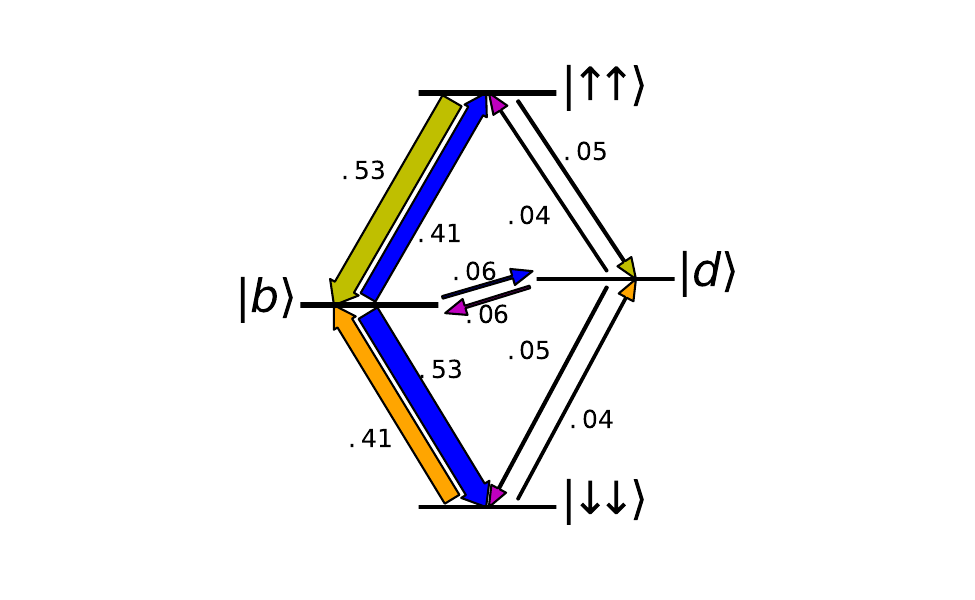}
}
\caption[]{\parbox[t]{140mm}{\raggedright
(\emph{left})
Energy levels of the coupled spin system as the coupling $\lambda_1$ is increased.
(The global shift of the energy levels has been kept in these data.)
Solid lines: no spin-orbit coupling $(\theta, \phi) = (0,0)$; dashed lines: $(\theta, \phi) = (0.37\,\pi, 0.42)$.
In the symmetric point $\lambda_1 = \lambda_2$, the dark state $|d\rangle$ touches the zero-energy line.
Parameters:
$\lambda_2 = 0.01$,
$\eV = 7.93\,\nu$, and $J_{\rm ex} \approx 0.13\,\nu$ at $\lambda_1 = \lambda_2$.
(\emph{right})
Dark and bright states formed by the effective interaction for the previous parameters
and $\lambda_1 = 1.26\,\lambda_2$. The thickness of the arrows and the given percentages represent the transition rates, normalized to the total rate $\approx 0.175\,\nu_1$ of transitions starting from the bright state (dark blue arrows). The transition $|b\rangle \leftrightarrow |d\rangle$ is due to dephasing processes, see Eq.(\ref{supp:eq_dark-bright-rate}). Note the weaker transitions involving the dark state $|d\rangle$.
}}
\label{suppfig:d-b-levels}
\end{figure}

\subsubsection{Two-spin relaxation processes}
\label{s:two-spin-jumps}

A diagonalization similar to the one performed for the Hamiltonian can also be applied to the array $(\gamma_{jk})$ that characterizes dissipative processes.
This results in a network of transitions between the system states that is visualised in Fig.\,\ref{suppfig:d-b-levels}(\emph{right}).

The diagonalization brings the remaining part of the master equation into the Lindblad form of Eq.\,(\ref{supp:e2350}). Due to the block-diagonal structure of~$(\gamma_{jk})$, we may consider first the subset listed in Eq.\,(\ref{supp:gamma-minus1-array}). This involves the operators $A_{1} = \tau_- \otimes \mathbbm{1}$ and $A_{2} = \mathbbm{1} \otimes \tau_-$ which are responsible for spin flips down the energy ladder. The `sandwich' terms, for example,  can be re-written as
\begin{equation}
\sum_{j,k \,=\, 1,2} \gamma_{-j,k} A_k^{\phantom\dagger} \rho A_j^\dagger
= \gamma_{-b} L_{-b}^{\phantom\dagger} \rho L_{-b}^\dagger
+ \gamma_{-d} L_{-d}^{\phantom\dagger} \rho L_{-d}^\dagger
\label{supp:eq_diagonalize-dissipation-terms}
\end{equation}
The transition rates $\gamma_{-b}$, $\gamma_{-d}$ are given by the eigenvalues of the hermitean $2\times2$ array in Eq.\,(\ref{supp:gamma-minus1-array})
\begin{eqnarray}
\begin{array}{c}
\gamma_{-b}\\
\gamma_{-d}
\end{array}
\bigg\}
&=& \half ( \eV + \nu )
\Big[
\half(\lambda_1 + \lambda_2)
\pm
\sqrt{
\lambda_{12}^2 \cos^2(\half\theta)
+
\tfrac{1}{4}(\lambda_1 - \lambda_2)^2
}
\,\Big]
\label{supp:eq_relaxation-rates-b-d}
\end{eqnarray}
We introduce another mixing angle (taking $0 \le \psi \le \pi$)
\begin{equation}
\tan\psi = \frac{ \lambda_{12} \cos(\half\theta) }{ \half( \lambda_1 - \lambda_2 ) }
\label{supp:eq_def-mixing-Ldown}
\end{equation}
and find that the jump operator $L_{-d}$ for the relaxation process at the rate $\gamma_{-d}$ has the composite form
\begin{equation}
L_{-d} =
- \sin(\psi/2) \eexp{i\phi/2} \tau_- \otimes \mathbbm{1}
+ \cos(\psi/2) \eexp{-i\phi/2} \mathbbm{1} \otimes \tau_-
\label{supp:eq_def-Ldown-bright}
\end{equation}
An `orthogonal' expression holds for $L_{-b}$. It is interesting to note that (i) the correlations among the bath spins at sites~1 and~2 produce a relaxation process where for example the excited state $|{\uparrow\uparrow}\rangle$ is mapped onto the entangled state
\begin{equation}
L_{-d} |{\uparrow\uparrow}\rangle
=
  \cos(\psi/2) \eexp{-i\phi/2} |{\uparrow\downarrow}\rangle
- \sin(\psi/2) \eexp{i\phi/2} |{\downarrow\uparrow}\rangle
\label{supp:eq_def-dark-state-from-down-jump}
\end{equation}
The second remark is (ii) that this state does not coincide in general with the eigenstate of the effective Hamiltonian ${\cal H}'$ [Eq.\,(\ref{supp:eigenstates})] because of the different expressions~(\ref{supp:eq_def-mixing-H}, \ref{supp:eq_def-mixing-Ldown}) for the mixing angles $\chi, \psi$. And finally, (iii) the state defined in Eq.\,(\ref{supp:eq_def-dark-state-from-down-jump}) is reached at the slower rate $\gamma_{-d} < \gamma_{-b}$ [$-$ sign in Eq.\,(\ref{supp:eq_relaxation-rates-b-d})], this is why we call this state `dark' rather than `bright'. A similar situation occurs in the collective fluorescence of two atoms at close distance where two superposition states with one atom excited, the other in the ground state, have either a small or a large emission rate. The slowly decaying (or dark) state favors the creation of (transient) entanglement by spontaneous emission.\cite{sm:Beige_2000, sm:Ficek_2003}

A similar analysis can be performed for the other sets of relaxation operators. From Eqs.\,(\ref{supp:gamma-plus1-array}), we find `bright' and `dark' excitation rates
\begin{eqnarray}
\begin{array}{c}
\gamma_{+b}\\
\gamma_{+d}
\end{array}
\bigg\}
&=&
\half ( \eV - \nu )
\Big[
\half(\lambda_1 + \lambda_2)
\pm
\sqrt{
\lambda_{12}^2 \cos^2(\half\theta)
+
\tfrac{1}{4}(\lambda_1 - \lambda_2)^2
}
\,\Big]
\label{supp:eq_excitation-rates-b-d}
\end{eqnarray}
They differ for near-degenerate spins only by a global scale factor $(\eV - \nu) / (\eV + \nu) = e^{-\beta^*\nu}$ from the relaxation rates $\gamma_{-b}$, $\gamma_{-d}$, while the mixing angle $\psi$ is the same. The corresponding jump operators are therefore just the hermitean conjugates of the previous ones, $L_{+d} = L_{-d}^\dagger$.
This does \emph{not} mean that jumping up from the ground state, the system reaches the same intermediate state as $L_{-d}| {\uparrow\uparrow} \rangle$ [Eq.\,(\ref{supp:eq_def-dark-state-from-down-jump})]. We rather get an exchange of the relative weights
\begin{equation}
L_{+d} | {\downarrow\downarrow} \rangle
=
- \sin(\psi/2) \eexp{-i\phi/2} |{\uparrow\downarrow}\rangle
+ \cos(\psi/2) \eexp{i\phi/2} |{\downarrow\uparrow}\rangle
\label{supp:eq_dark-state-from-jumping-up}
\end{equation}
This coincides with Eq.\,(\ref{supp:eq_def-dark-state-from-down-jump}) only in the symmetric case $\psi = \pi/2$. See Sec.\,\ref{supp:resonant-spins} for more details.

The transitions generated by the Lindblad operators are represented by the colored arrows in Fig.\ref{suppfig:d-b-levels}(\emph{right}), their thickness being proportional to the corresponding rate.
We conjecture that no stationary entanglement can be generated in this way: the populations of the dark and bright states will equilibrate to the same value because the up and down rates are in the same ratio $e^{-\beta^*\nu}$. As far as the bath-induced relaxation processes are concerned, the two states appear to reach the same Boltzmann weights. This will be checked by an explicit calculation in Sec.\,\ref{s:steady-states} taking into account the interplay between the effective Hamiltonian ${\cal H}'$ and the jump processes.

It remains to consider the dephasing processes determined by the coefficients $\gamma_{zz}, \ldots \gamma_{z'z'}$ in Eq.(\ref{supp:gamma-zz-array}). One might again compute eigenvalues and introduce fast and slow dephasing rates, the analysis is simpler in the bare basis, however.
Consider an eigenstate $|ab\rangle$ with $a, b = {\uparrow}, \downarrow$ of the bare Hamiltonian ${\cal H}_0$. The dephasing Lindblad operators are generated by the single-spin operators $A_z = \tau_z \otimes \mathbbm{1}$ and $A_{z'} = \mathbbm{1} \otimes \tau_z$ of which $|ab\rangle$ is an eigenstate.
By writing the Lindblad terms in the master equation in the form (the conjugation is actually not needed here)
\begin{equation}
  A_k^{\phantom\dagger} \rho A_j^\dagger
- \tfrac{1}{2} A_j^\dagger A_k^{\phantom\dagger} \rho
- \tfrac{1}{2} \rho A_j^\dagger A_k^{\phantom\dagger}
=
  \tfrac{1}{2} \big[ A_k^{\phantom\dagger} \rho, \, A_j^\dagger \big]
+ \tfrac{1}{2} \big[ A_k^{\phantom\dagger}, \, \rho A_j^\dagger \big]
\label{supp:eq_re-write Lindblad}
\end{equation}
it is simple to see that the diagonal parts of the density operator like
$|ab\rangle\langle ab|$ are not affected by dephasing (they are mapped to zero). But coherences (off-diagonal elements) $|ab\rangle\langle cd|$ with $(a,b) \ne (c,d)$ are affected; it turns out that they are mapped to a multiple of $|ab\rangle\langle cd|$. The (negative) prefactors determine the dephasing rate of this coherence (in the eigenbasis of ${\cal H}_0$). The results are collected in the following Table (void entries are zero).
\begin{align}{\label{supp:eq_dephasing-rates}}
\begin{array}[c]{c|cccc|}
ab \backslash cd
	& \uparrow\uparrow & \uparrow\downarrow & \downarrow\uparrow & \downarrow\downarrow
\\
\hline
\uparrow\uparrow
&  & 2 \gamma_{z'z'} & 2 \gamma_{zz}
		& 2 \gamma_{\rm p}
\\
\uparrow\downarrow
& \cdot  & & 2 \gamma_{\rm a} & 2 \gamma_{zz}
\\
\downarrow\uparrow
&  \cdot  &  \cdot  & & 2 \gamma_{z'z'}
\\
\downarrow\downarrow
& \cdot  & \cdot   & \cdot   &
\\
\hline
\end{array}
\qquad
\begin{array}{r}
\gamma_{\rm p}
\\
\gamma_{\rm a}
\end{array}
\bigg\} &=
\gamma_{zz} + \gamma_{z'z'} \pm ( \gamma_{zz'} + \gamma_{z'z} )
\\[-6.5ex]
&= \frac{ \eV }{ 4 } \big(
\lambda_1 + \lambda_2 \pm 2 \lambda_{12} \cos(\half\theta) \cos\phi
\big)
\nonumber
\end{align}
For simplicity, we did not write out the lower triangle of entries (marked with $\cdot$) which is symmetric to the upper one. The off-diagonal coefficients $\gamma_{zz'}$ and $\gamma_{z'z}$ thus increase the dephasing rate $\gamma_{\rm p}$ for superpositions of `parallel' states $|{\uparrow\uparrow}\rangle$ and $|{\downarrow\downarrow}\rangle$, while the antiparallel states decohere at the slower rate $\gamma_{\rm a}$. This is similar to the selection rules for two nearby spins excited by a homogeneous field.

As is well-known, dephasing speeds up the decay of the coherence among any pair of states. For the entangled states defined in Eq.(\ref{supp:eigenstates}), when we applying this part of the master equation to $| d \rangle \langle b |$ and projecting onto the same matrix element, the contribution to the decoherence rate is
\begin{equation}
- \tr\Big\{ | b \rangle \langle d |
\sum_{j,k = z,z'} \gamma_{jk}
\big(
A_k | d \rangle \langle b | A_j
- \tfrac{1}{2}
A_j A_k | d \rangle \langle b |
- \tfrac{1}{2}
| d \rangle \langle b | A_j A_k
\big)
\Big\}
= \gamma_{\rm a} ( 1 + \cos^2\chi )
\label{supp:eq-bright-dark-decoherence}
\end{equation}
as a straightforward calculation shows. In this basis, it turns out that
dephasing also leads to transitions among different eigenstates of the effective Hamiltonian ${\cal H}'$. When the corresponding Lindblad operators act on the diagonal element $|d\rangle \langle d|$, for example, one finds that, among other terms, a projector $|b\rangle \langle b|$ appears. The rate for this transition $d \to b$ we calculate as
\begin{equation}
r_{d \to b} =
\sum_{j,k = z,z'} \gamma_{jk}
\langle b | A_k | d \rangle \langle d | A_j | b \rangle
= \gamma_{\rm a} \sin^2\chi
\label{supp:eq_dark-bright-rate}
\end{equation}
where $\chi$ was defined in Eq.(\ref{supp:eq_def-mixing-H}).
(The terms with the ordering $A_j A_k | d \rangle \langle d |$ do not contribute to $r_{d \to b}$ because $|d\rangle$ and $|b\rangle$ are orthogonal. Lindblad operators that jump up or down cannot contribute neither.)

We may also consider the dissipative dynamics of spin-spin correlations by expanding the density operator in products of spin operators,
$\rho = \sum_{\alpha, \beta} \rho_{\alpha\beta} \tau_{\alpha} \otimes \tau_{\beta}$. By taking $\alpha, \beta = 0, z, \pm$ with $\tau_{0} = \mathbbm{1}$, this generalizes the expansion in terms of the Bloch vector (see Sec.\,\ref{supp:sec-Bloch-equations} and Ref.\,\onlinecite{sm:shnirman1}). The trace of $\rho$ is normalized if we take $\rho_{00} = \quarter$. The coefficients $\rho_{0z}$, $\rho_{z0}$ and $\rho_{zz}$ determine the populations in the ${\cal H}_0$ eigenbasis, while any element with at least one index $\alpha$ or $\beta = \pm$ describes coherences in this basis. The latter matrix elements are generally complex and satisfy, for example, $\rho_{+z}^* = \rho_{-z}$.
The dephasing processes proportional to $\gamma_{zz}$ and $\gamma_{z'z'}$ can be worked out similar to the single-spin case of Eqs.(\ref{supp:e2501}) before.
For one of the two `mixed terms', we find
\beq{33}
&&\half\gamma_{zz'} \big\{
A_{z'} \rho A_z - A_z A_{z'} \rho + \text{h.c.}
\big\}
\nonumber
\\
&& = \half\gamma_{zz'} \sum_{\alpha,\beta}\rho_{\alpha\beta}
\left[
(\mathbbm{1} \otimes\tau_z)(\tau_\alpha\otimes\tau_\beta) (\tau_z\otimes \mathbbm{1})
-(\tau_z\otimes \mathbbm{1}) (\mathbbm{1} \otimes \tau_z) (\tau_\alpha\otimes\tau_\beta)
\right]
+ \text{h.c.}
\nonumber
\\
&&
= \half\gamma_{zz'} \sum_{\alpha,\beta}\rho_{\alpha\beta}
[\tau_\alpha, \, \tau_z]\otimes\tau_z\tau_\beta + \text{h.c.}
\nonumber
\\
&& =
 \half\gamma_{zz'} \sum_{a=\pm,\beta}\rho_{a\beta}(-2a\tau_{a})\otimes \tau_z\tau_\beta
+ \text{h.c.}
\nonumber
\\
&&
= \gamma_{zz'} \sum_{a=\pm,\beta=0,z}
\left\{
  \rho_{a\beta}(-a\tau_a)\otimes\tau_z\tau_\beta
+ \rho_{-a,\beta}(-a\tau_{-a})\otimes\tau_\beta\tau_z
\right\}
\nonumber
\\
&&
\hphantom{{} =}
{} +
\gamma_{zz'} \sum_{a=\pm,b=\pm}
\left\{
  \rho_{ab}(-a\tau_a)\otimes(b\tau_b)
+ \rho_{-a,-b}(-a\tau_{-a})\otimes (b\tau_{-b})
\right\}
\nonumber
\\
&&
= 2\gamma_{zz'} \sum_{a=\pm,b=\pm}\rho_{ab}(-a\tau_a)\otimes(b\tau_b)
\eeq
On the third line above, the commutator only contributes for $\alpha = a = \pm$,
and two lines down, the terms with $\beta = 0, z$ cancel with their h.c.
Similarly, in the other mixed term with indices ${z'z}$, only $\beta=\pm$ contributes and its calculation yields
\beq{34}
&& \half\gamma_{z'z} \big\{ A_{z}\rho A_{z'} - A_{z'} A_{z}\rho + \text{h.c.} \big\}
=
2\gamma_{z'z}  \sum_{a=\pm,b=\pm}\rho_{ab}(a\tau_a)\otimes(-b\tau_b)
\eeq
which is the same as Eq.(\ref{supp:e33}) and only affects coherences. We check that this is consistent with Eq.(\ref{supp:eq_dephasing-rates}): In the basis $( | {\uparrow\uparrow} \rangle, | {\uparrow\downarrow} \rangle, \ldots )$ used in the Table there, the operator identity $\tau_+ \otimes \tau_- = | {\uparrow\downarrow} \rangle \langle {\downarrow\uparrow} |$ holds. The term $-2(\gamma_{zz'} + \gamma_{z'z})$ then corresponds to the (negative) coefficient of $\tau_+ \otimes \tau_-$ in the sum of Eqs.(\ref{supp:e33}, \ref{supp:e34}).

\begin{table}[bth]
$$
\begin{array}[t]{l|l}
& \text{Couplings}
\\ \hline
\lambda_j 	& 16\pi J_j^2 N(\epsilon_F)^2
\\
\lambda_{12}& 16\pi J_1 J_2 N(\epsilon_F)^2
\\
\delta 	& \nu \ln(\eV/\Lambda)
\\[0.5ex]
\hline
\end{array}
\hspace*{12mm}
\begin{array}[t]{l|l}
& \text{Effective Hamiltonian}
\\
\hline
\nu_j' 		& \nu - \frac{1}{2\pi} \lambda_j \delta
\\
J_{\rm ex} 	& \tfrac{1}{4\pi} \lambda_{12} \deltaTwo \cos\half\theta \cos\phi
\\
J' & \tfrac{1}{2\pi} \lambda_{12} \deltaTwo \cos\half\theta
\\[0.5ex]
\hline
\end{array}
\hspace*{12mm}
\begin{array}[t]{l|l}
& \text{Transition rates}
\\
\hline
\Gamma_j 	& \lambda_j \eV
\\
\gamma_j 	& \lambda_j \nu
\\
G			& \lambda_{12} \eV \cos\half\theta
\\
g			& \lambda_{12} \nu \cos\half\theta
\\[0.5ex]
\hline
\end{array}
$$
\caption[]{\parbox[t]{140mm}{\raggedright
Notation used in Eqs.\,(\ref{supp:e6410}--\ref{supp:e6410c}) and Eqs.\,(\ref{supp:e6420}--\ref{supp:e6420c}).}}
\label{t:dictionary-gamma-delta}
\end{table}

\subsubsection{Master equation}
\label{s:master-equation}

Now that all building blocks of the master equation are put together, we are ready to write down the equations of motion for the elements of the density matrix. We use the expansion coefficients $\rho_{\alpha\beta}$ introduced after Eq.\,(\ref{supp:eq-bright-dark-decoherence}), and for the ease of presentation, we display separately the bath-induced Hamiltonian and the dissipative terms. Abbreviations are collected in Table~\ref{t:dictionary-gamma-delta}.

The contribution $-i[{\cal H}_0' + {\cal H}', \rho]$ of the Hamiltonian (free plus exchange and DM interactions) is
\begin{eqnarray}
\frac{ d \rho_{zz} }{ d t }\bigg|_{\cal H} &=& 0
\nonumber\\
\frac{ d \rho_{z0} }{ d t }\bigg|_{\cal H} &=& -\tfrac{i}{2} J' \big( {\rm e}^{i\phi}  \rho_{+-} - {\rm e}^{-i\phi}  \rho_{-+} \big)
\nonumber\\
\frac{ d \rho_{0z} }{ d t }\bigg|_{\cal H} &=& + \tfrac{i}{2} J' \big( {\rm e}^{i\phi}  \rho_{+-} - {\rm e}^{-i\phi}  \rho_{-+} \big)
\nonumber\\
\frac{ d \rho_{+-} }{ d t }\bigg|_{\cal H} &=& -i (\nu_1'-\nu_2')  \rho_{+-}
- 2 i J' \,{\rm e}^{-i\phi} \big(  \rho_{z0} - \rho_{0z} \big)
\label{supp:e6410}
\end{eqnarray}
\begin{eqnarray}
\frac{ d \rho_{+0} }{ d t }\bigg|_{\cal H} &=& -i \nu_1'  \rho_{+0} + 2 i J_{\rm ex}  \rho_{+z} -i J' \,{\rm e}^{-i\phi}  \rho_{z+}
\nonumber\\
\frac{ d \rho_{+z} }{ d t }\bigg|_{\cal H} &=& -i \nu_1'  \rho_{+z} + 2 i J_{\rm ex}  \rho_{+0} -i J' \,{\rm e}^{-i\phi}  \rho_{0+}
\nonumber\\
\frac{ d \rho_{0+} }{ d t }\bigg|_{\cal H} &=& -i \nu_2'  \rho_{0+} + 2 i J_{\rm ex}  \rho_{z+} -i J' \,{\rm e}^{i\phi}  \rho_{+z}
\nonumber\\
\frac{ d \rho_{z+} }{ d t }\bigg|_{\cal H} &=& -i \nu_2'  \rho_{z+} + 2 i J_{\rm ex}  \rho_{0+} -i J' \,{\rm e}^{i\phi}  \rho_{+0}
\label{supp:e6410b}
\end{eqnarray}
\begin{eqnarray}
\frac{ d \rho_{++} }{ d t }\bigg|_{\cal H} &=& -i (\nu_1'+\nu_2')  \rho_{++}
\label{supp:e6410c}
\end{eqnarray}
Equations that arise by simple conjugation have been suppressed, for example, $\rho_{-+} = \rho_{+-}^*$.
We have grouped the elements into blocks that are coupled only among themselves.
The dissipative parts of the master equation give the additional terms (index $L$ for `Lindblad'):
\begin{eqnarray}
\frac{ d\rho_{zz} }{ d t }\bigg|_{L} &=& - (\Gamma_1 + \Gamma_2) \rho_{zz} - \gamma_2 \rho_{z0} - \gamma_1 \rho_{0z} + \half G \big( {\rm e}^{i\phi}  \rho_{+-} + {\rm e}^{-i\phi}  \rho_{-+} \big)
\nonumber
\\
\frac{ d\rho_{z0} }{ d t }\bigg|_{L} &=& - \Gamma_1  \rho_{z0} - \gamma_1 \rho_{00}-\quarter g \big( {\rm e}^{i\phi}  \rho_{+-} + {\rm e}^{-i\phi}  \rho_{-+} \big)
\nonumber
\\
\frac{ d\rho_{0z} }{ d t }\bigg|_{L} &=& - \Gamma_2 \rho_{0z} - \gamma_2 \rho_{00} -\quarter g \big( {\rm e}^{i\phi}  \rho_{+-} + {\rm e}^{-i\phi}  \rho_{-+} \big)
\nonumber
\\
\frac{ d\rho_{+-} }{ d t }\bigg|_{L} &=& -\big[ \Gamma_1 + \Gamma_2 - G \cos\phi \big]  \rho_{+-} + g \,{\rm e}^{-i\phi} \big( \rho_{z0} + \rho_{0z} \big) + 2  G \,{\rm e}^{-i\phi}  \rho_{zz}
\label{supp:e6420}
\end{eqnarray}
\begin{eqnarray}
\frac{ d\rho_{+0} }{ d t }\bigg|_{L} &=& - \Gamma_1  \rho_{+0} + \half g \,{\rm e}^{-i\phi}  \rho_{z+}
\nonumber
\\
\frac{ d\rho_{+z} }{ d t }\bigg|_{L} &=& - (\Gamma_1 + \Gamma_2) \rho_{+z} - \gamma_2 \rho_{+0} -\half  g \,{\rm e}^{-i\phi}  \rho_{0+}- G \,{\rm e}^{-i\phi}  \rho_{z+}
\nonumber
\\
\frac{ d\rho_{0+} }{ d t }\bigg|_{L} &=& - \Gamma_2 \rho_{0+} + \half g \,{\rm e}^{i\phi}  \rho_{+z}
\nonumber
\\
\frac{ d\rho_{z+} }{ d t }\bigg|_{L} &=& - (\Gamma_1 + \Gamma_2) \rho_{z+} - \gamma_1 \rho_{0+} -\half  g \,{\rm e}^{i\phi}  \rho_{+0}- G \,{\rm e}^{i\phi}  \rho_{+z}
\label{supp:e6420b}
\end{eqnarray}
\begin{eqnarray}
\frac{ d\rho_{++} }{ d t }\bigg|_{L} &=& -\big[ \Gamma_1 + \Gamma_2 + G \cos\phi \big]  \rho_{++}
\label{supp:e6420c}
\end{eqnarray}
This set of equations is consistent with the blocks of the Hamiltonian terms above.
We recall that $\rho_{00} = \quarter$ remains constant.

\subsubsection{Stationary state}
\label{s:steady-states}

The stationary state of the two spins contains the dark state under certain conditions that we derive here in a simple way. The block structure helps in the calculation, for those blocks that do not involve the coefficient $\rho_{00}$ relax to zero in the stationary state, their equations of motion being linear and homogeneous. Only the coefficients $\rho_{zz}$, $\rho_{z0}$, $\rho_{0z}$, $\rho_{+-}$, and $\rho_{-+}$ need to be calculated.
Splitting
\begin{equation}
{\rm e}^{i\phi} \rho_{+-} = \half ( x + i y )
\label{supp:e-def-xy}
\end{equation}
into real and imaginary parts, we get
\begin{eqnarray}
\rho_{zz} &=& \frac{ 1 }{ \Gamma_1 + \Gamma_2 } \big( \half G x - \gamma_2 \rho_{z0} - \gamma_1 \rho_{0z} \big)
\nonumber\\
\rho_{z0} &=& - \frac{ 1 }{ 2 \Gamma_1 } \big( \half \gamma_1 + \half g x - J' y \big)
\nonumber\\
\rho_{0z} &=& - \frac{ 1 }{ 2 \Gamma_2 } \big( \half \gamma_2 + \half g x + J' y \big)
\label{supp:e-diagonal-populations}
\end{eqnarray}
Insert this into the expression for the stationary value of ${\rm e}^{i\phi} \rho_{+-}$ from Eqs.\,(\ref{supp:e6410}, \ref{supp:e6420})
\begin{equation}
{\rm e}^{i\phi} \rho_{+-} =
\frac{ - 2 i J' (\rho_{z0} - \rho_{0z}) + g (\rho_{z0} + \rho_{0z}) + 2 G \rho_{zz}
}{ \Gamma_1 + \Gamma_2 - G \cos\phi + i (\nu_1' - \nu_2') }
\label{supp:e-expression-rho+-}
\end{equation}
and collect in the numerator the inhomogeneous terms on the right-hand side (that arise from $\rho_{00} = \quarter$):
\begin{equation}
\frac{i J'}{2} \Big( \frac{ \gamma_1 }{ \Gamma_1 } - \frac{ \gamma_2 }{ \Gamma_2 } \Big)
- \frac{ g }{ 4 } \Big( \frac{ \gamma_1 }{ \Gamma_1 } + \frac{ \gamma_2 }{ \Gamma_2 } \Big)
+ \frac{ G }{ 2 } \frac{ \gamma_1 \gamma_2 }{ \Gamma_1 + \Gamma_2 }
\Big( \frac{ 1 }{ \Gamma_1 } + \frac{ 1 }{ \Gamma_2 } \Big)
\end{equation}
This vanishes because by definition, the identities $\gamma_1 / \Gamma_1 = \gamma_2 / \Gamma_2 = g / G$ hold; the second and third terms mutually cancel.
It thus seems that ${\rm e}^{i\phi} \rho_{+-}$ actually satisfies a homogeneous equation, and we may expect it to vanish.

To simplify the following calculations, we focus on the special case where $\Gamma_1 = \Gamma_2 \equiv \Gamma$, $\gamma_1 = \gamma_2 \equiv \gamma$, and $G = \Gamma \cos(\half\theta)$. Collecting the coefficients of $x$ and $y$, the expression~(\ref{supp:e-expression-rho+-}) becomes
\begin{equation}
\frac{ x + iy }{ 2 } = \frac{ \half G^2 x - 2 i J^{\prime 2} y
}{ \Gamma \big[ 2 \Gamma - \Gamma \cos(\half\theta) \cos\phi + i (\nu_1' - \nu_2') \big] }
\label{supp:e-x+iy}
\end{equation}
Separating into real and imaginary parts, we find a linear system whose determinant is proportional to
\begin{equation}
\det
= \quarter \big( 2 \Gamma - \Gamma \cos(\half\theta) \cos\phi - \Gamma \cos^2(\half\theta)\big)
\big( \Gamma + 4 J^{\prime 2} / \Gamma \big) + \quarter (\nu_1' - \nu_2')^2
\label{supp:e-}
\end{equation}
Because $2 - \cos(\half\theta) \cos\phi \ge 1$ and $\cos^2(\half\theta) \le 1$, this determinant is a sum of non-negative terms and can only vanish in the
\begin{equation}
\text{symmetric point}: \quad
\lambda_1 = \lambda_2
\quad \text{and} \quad
(\theta, \phi) = (0,0)
\label{supp:e-symmetric-conditions}
\end{equation}
(see Table\,\ref{t:dictionary-gamma-delta}).
This is the situation we study in more detail in Sec.\,\ref{supp:resonant-spins} below.
Away from this special point, we conclude that $x = y = 0$, hence $\rho_{+-} = 0$ in the stationary state. Since this coefficient is proportional to the coherence between the intermediate basis states $|{\uparrow\downarrow}\rangle$ and $|{\downarrow\uparrow}\rangle$, there is no stationary entanglement.

From Eqs.\,(\ref{supp:e-diagonal-populations}), we get for the nonzero elements of the stationary density matrix
\begin{eqnarray}
\rho_{\rm st}: \qquad
\rho_{z0} &=& - \frac{ \gamma_1 }{ 4 \Gamma_1 }
\,,\qquad
\rho_{0z} = - \frac{ \gamma_2 }{ 4 \Gamma_2 }
\,,\qquad
\rho_{zz} = \frac{ \gamma_1 \gamma_2 }{ 4 \Gamma_1 \Gamma_2 }
\label{supp:e-result-diagonal-populations}
\end{eqnarray}
This yields the following spin expectation values
\begin{eqnarray}
\langle \tau_z\otimes \mathbbm{1}\rangle_{\rm st} &=& 4\rho_{z0}
= -\frac{\nu}{\eV}
=
\langle \mathbbm{1} \otimes \tau_z\rangle_{\rm st}
\nonumber\\
\langle \tau_z\otimes\tau_z\rangle_{\rm st} &=& 4\rho_{zz}
=
\langle \tau_z\otimes \mathbbm{1}\rangle_{\rm st}
\langle \mathbbm{1} \otimes \tau_z\rangle_{\rm st}
\label{supp:e-result-no-spin-z-correlations}
\end{eqnarray}
There are no correlations between the spins, and using the effective temperature $1/\beta^*$, each of the spins has the familiar longitudinal average $\langle \tau_z \otimes \mathbbm{1} \rangle_{\rm st}
= -\tanh(\half\beta_1^*\nu)$.

At this order in our approximations, the level shifts due to the effective interaction do not modify the equilibrium populations.
This would have been different if an exchange interaction had been included right from the start in the system Hamiltonian, before evaluating the system-bath interaction in second-order perturbation theory (see Ref.\,\onlinecite{sm:shavit} and Sec.\,\ref{sm:dd-interaction} below).

\subsection{Spin entanglement at the symmetric point}
\label{supp:resonant-spins}

\subsubsection{Decoupling of the dark state}

To illustrate the significance of the symmetric point, consider first the equation of motion for the dark state population $p_{d} = \langle d | \rho | d \rangle$ [see Eq.\,(\ref{supp:eigenstates})]. Re-writing the projector $|d\rangle\langle d|$ in terms of Pauli matrices, we find
\begin{equation}
\frac{ d p_d }{ dt } =
- \frac{ d \rho_{zz} }{ dt }
+ \cos\chi
\Big(
    \frac{ d \rho_{z0} }{ dt }
  + \frac{ d \rho_{0z} }{ dt }
\Big)
- \frac{ \sin\chi }{ 2 }
\Big(
    {\rm e}^{{\rm i} \phi} \frac{ d \rho_{+-} }{ dt }
  + {\rm e}^{-{\rm i} \phi} \frac{ d \rho_{-+} }{ dt }
\Big)
\label{supp:eqn-motion-dark-state}
\end{equation}
where $\chi$ was defined in Eq.\,(\ref{supp:eq_def-mixing-H}).
This generalizes Eq.(\ref{e08}) of the main text to parameters beyond the symmetric point. We insert the equations of motion~(\ref{supp:e6410}) and~(\ref{supp:e6420}) and find
\begin{eqnarray}
\frac{ d p_d }{ dt } &=&
- i J' \cos\chi \big(
    {\rm e}^{{\rm i} \phi} \rho_{+-}
  - {\rm e}^{-{\rm i} \phi} \rho_{-+}
  \big)
+ \frac{ i }{ 2 } (\nu_1' - \nu_2') \sin\chi \big(
    {\rm e}^{{\rm i} \phi} \rho_{+-}
  - {\rm e}^{-{\rm i} \phi} \rho_{-+}
  \big)
\nonumber
\\
&& {}
+ (\gamma_2 - \gamma_1) \cos(\chi) \rho_{00}
+ (\Gamma_1 + \Gamma_2 - 2 G \sin\chi) \rho_{zz}
\nonumber
\\
&& {}
+ (\gamma_2 - \Gamma_1 \cos\chi - g \sin\chi) \rho_{z0}
+ (\gamma_1 - \Gamma_2 \cos\chi - g \sin\chi) \rho_{0z}
\nonumber
\\
&& {}
+ \half \big[
(\Gamma_1 + \Gamma_2) \sin\chi - G (1 + \sin\chi \cos\phi)
\big]
\big(
    {\rm e}^{{\rm i} \phi} \rho_{+-}
  + {\rm e}^{-{\rm i} \phi} \rho_{-+}
  \big)
\label{supp:eqn-motion-pd-2}
\end{eqnarray}
Requiring that this vanishes for any initial values of $\rho_{\alpha\beta}$ provides an alternative way of identifying the symmetric point.
The mixing angle $\chi$ is such that the first line vanishes.
The other rates cancel only when the couplings $\lambda_j$ are equal and spin-orbit coupling is absent so that $\chi = \pi/2$, $\gamma_j = g$ and $\Gamma_j = G$ ($j = 1,2$, see Table\,\ref{t:dictionary-gamma-delta}).

What actually happens in the symmetric case, is that the dark state ``decouples'' from the other three states. As mentioned earlier, the mixing angles relevant for the Hamiltonian ($\chi$) and the relaxation rates ($\psi$) take both the value $\pi/2$ at the symmetric point so that the jump operators $L_{d}$ and $L_{d}^\dagger$ in Sec.\,\ref{s:two-spin-jumps} reach the dark state: $L_d |{\uparrow\uparrow}\rangle = |d\rangle = -L_d^\dag |{\downarrow\downarrow}\rangle$. The corresponding rates $\gamma_{\pm d}$, however, vanish when we set $\lambda_1 = \lambda_2 = \lambda_{12}$ and $\theta = 0$ in Eqs.\,(\ref{supp:eq_relaxation-rates-b-d}, \ref{supp:eq_excitation-rates-b-d}). The same is true for the dephasing rate $\gamma_{\rm a}$ in the subspace spanned by $|{\uparrow\downarrow}\rangle$, $|{\uparrow\downarrow}\rangle$ [Eq.\,(\ref{supp:eq_dephasing-rates})]. This is illustrated by comparing the Figures\,\ref{suppfig:d-b-levels}\,(right) and~\ref{fig:symmetric-network-of-rates}\,(left).

This behaviour could have been guessed from the structure of the system-bath interaction. At the symmetric point, the spin of the bath electrons in Eq.\,(\ref{supp:e28}) couples to the total spin ${\bf S}$ of the two tunnelling sites:~\cite{sm:bh}
\begin{equation}
{\cal H}_{SE} =
J c_R^\dagger\bfsigma c_L^{\phantom\dagger} \cdot
\big(
\bftau \otimes \mathbbm{1}
+ \mathbbm{1} \otimes\bftau
\big)
+ \text{h.c.} =
J c_R^\dagger\bfsigma c_L^{\phantom\dagger} \cdot {\bf S}
+
J c_L^\dagger\bfsigma c_R^{\phantom\dagger} \cdot {\bf S}
\label{supp:e-Hamiltonian-SE-with-exchange-symmetry}
\end{equation}
The dark state $|d\rangle$ coincides in the symmetric point with the familiar singlet state: its total spin is $S = 0$ and for all spin components ${\bf S}|d\rangle = 0$. We have checked that the energy of the dark state is not shifted by the contact with the bath: it is actually the global shift $-(3/4\pi)\lambda \Delta$ that compensates for the eigenvalue $E_d = J_{\rm ex} + J'$ [Eq.\,(\ref{supp:eigenstates})] of the dark state. This can be seen in Fig.\,\ref{suppfig:d-b-levels}(\emph{left}) for $\lambda_1 = \lambda_2$.

\subsubsection{Family of stationary states}

The singlet-triplet decoupling also explains why the stationary state is not unique, but rather a one-dimensional manifold parametrized by the coefficient $x = 2\mathop{\rm Re}\rho_{+-}$ from Eqs.\,(\ref{supp:e-def-xy}, \ref{supp:e-diagonal-populations}). [It follows from the imaginary part of Eq.\,(\ref{supp:e-x+iy}) that $y = 0$ because its coefficients on both sides have opposite signs.] The weight of the dark state is set by the initial conditions, while the triplet sector of the system state relaxes to a density operator $\rho_{\rm tr}$ with populations on the states $|{\uparrow\uparrow}\rangle$, $|b\rangle$, and $|{\downarrow\downarrow}\rangle$ only.

The symmetric point is thus hosting a family of stationary states that `mix' between $\rho_{\rm tr}$ and the pure singlet state $|d\rangle$. The nonzero coefficients of the Pauli matrix expansion within this family are from Eqs.\,(\ref{supp:e-diagonal-populations}):
\begin{equation}
\rho_{0z} = - \frac{ \gamma }{ 4\Gamma } (1 + x) = \rho_{z0}
,\qquad
\rho_{zz} = \frac{ x }{ 4 } + \frac{ \gamma^2 }{ 4\Gamma^2 } (1 + x)
,\qquad
\rho_{+-} = \frac{ x }{ 2 } = \rho_{-+}
\label{supp:e-triplet-mixture}
\end{equation}
The correlations among the spin components are now
\begin{eqnarray}
\langle \tau_z \otimes \tau_z \rangle -
\langle \tau_z \otimes \mathbbm{1} \rangle \langle \mathbbm{1} \otimes \tau_z \rangle
&=&
x - \frac{ \gamma^2 }{ \Gamma^2 } x (1+x)
\nonumber
\\
\text{and} \qquad
\langle \tau_x \otimes \tau_x \rangle &=& \langle \tau_y \otimes \tau_y \rangle = x
\label{supp:e-spin-spin-correlations}
\end{eqnarray}
The diagonal state $\rho_{\rm st}$ found in Sec.\,\ref{s:steady-states} arises for the special value $x = 0$ (no correlations).
The range of the parameter $x$ is limited by the requirement that $\rho$ be a physical state: its eigenvalues must be positive. A brief calculation gives the constraints $-1 \le x \le x_{\rm tr} = (1 -  \gamma^2 / \Gamma^2)/(3 + \gamma^2 / \Gamma^2)$. The endpoints of this interval correspond to the dark state $|d\rangle$ and the triplet state $\rho_{\rm tr}$ mentioned before, respectively. In the entire range, the spin correlations are nearly isotropic because of the small ratio $\gamma/\Gamma = \nu/\eV$ in Eq.\,(\ref{supp:e-spin-spin-correlations}).

From the ratio of populations of the triplet states $|{\uparrow\uparrow}\rangle$, $|b\rangle$, and $|{\downarrow\downarrow}\rangle$, we compute a `Boltzmann factor' and find $(\Gamma - \gamma)/(\Gamma + \gamma)$. This coincides with the single-spin value $(\eV - \nu)/(\eV + \nu)$ found earlier [Eq.\,(\ref{supp:e20})], for all $x$.
In particular the triplet equilibrium state $\rho_{\rm tr}$ is characterized by the same effective temperature $1/\beta^*$ as away from the symmetric point.

We now evaluate the steady-state entanglement in the one-parameter manifold of steady states. It is convenient to re-parametrize it in terms of a mixing probability $p \in [0,1]$ between, on the one hand,
the equilibrated triplet $\rho_{\rm tr}$ found before
and the dark state on the other: $\rho(p) = (1 - p) \rho_{\rm tr} + p\, |d\rangle\langle d|$. From the results above, we find simply $p = 1 -  (x + 1)/(x_{\rm tr} + 1)$. This mixture is similar to the well-known example of a Werner state where a maximally entangled component like $|d\rangle\langle d|$ is combined with a totally mixed state. \cite{sm:Werner_1989} (In our case however, $\rho_{\rm tr}$ shows some correlations, as seen above.)

The correlations between the two spins can be characterized by the correlation entropy $S_{12}$ (also known as quantum mutual information) defined by
\begin{equation}
S_{12} = S(\rho_1) + S(\rho_2) - S( \rho )
\label{supp:e-def-S12}
\end{equation}
where $S( \rho )$ is the von Neumann entropy: in terms of the eigenvalues $\{ p_i \}$ of $\rho$, we have $S(\rho) = - \sum_{i} p_i \log p_i$. The state $\rho_1$ is the reduced state of spin~1, obtained by tracing out the other spin; its Pauli matrix expansion is $\rho_1 = 2 \big( \rho_{00}\mathbbm{1} + \rho_{z0} \tau_z + \rho_{+0} \tau_{+} + \rho_{-0} \tau_{-} \big)$. (Analogously for $\rho_2$.) In a maximally entangled state, $S_{12} = 2 \log 2$ because $S( \rho ) = 0$ (a pure state has zero entropy), while the partial states $\rho_{1}$ and $\rho_{2}$ are maximally mixed, i.e., $\rho_1 = \half\mathbbm{1}$. Figure~\ref{dark}\,(right) of the main paper shows that the correlation entropy is nonzero for nearly all mixing parameters. The only exception is the diagonal state $\rho_{\rm st}$ equal to the unique stationary state when we leave the symmetric point (found in Sec.\,\ref{s:steady-states} above).

The correlation entropy detects correlations, but is not specific to entanglement. Two measures can be computed easily from the system density matrix: the entanglement of formation ${\cal E}_F$ (introduced by Wootters~\cite{sm:Wootters_1998} and related to the concurrence) and the logarithmic negativity ${\cal E}_N$ introduced by Vidal and Werner~\cite{sm:Vidal_2002} based on work by Peres~\cite{sm:Peres_1996} and the Horodecki family~\cite{sm:Horodecki_1996}. They are constructed to give zero for separable (non-entangled) states, be they pure or mixed. Their definitions:

\paragraph*{Entanglement of formation ${\cal E}_F$.} From the two-spin density operator $\rho$, construct, in the eigenbasis of ${\cal H}_0$, the operator $\tilde\rho = (\tau_y \otimes \tau_y) \rho^* (\tau_y \otimes \tau_y)$ where the matrix elements of $\rho^*$ are complex conjugate to those of $\rho$. (Using the expansion into Pauli matrices $\tau_\alpha \otimes \tau_\beta$, the map $\rho \mapsto \tilde\rho$ flips the signs of the coefficients $\rho_{\alpha 0}$ and $\rho_{0\beta}$ for $\alpha, \beta = z, \pm$, as if the single-particle spin operators had changed sign.) Compute the eigenvalues of the non-hermitean matrix $\tilde\rho\,\rho$ (they are real and non-negative) and denote $\{ r_i \}$ their square roots. Sort these so that $r_1 > r_2 > \ldots$, construct the concurrence $C = \max(0, r_1 - r_2 - r_3 - r_4)$ and $r = \half(1 + \sqrt{1 - C^2})$. Then we have $0 \le r \le 1$ and ${\cal E}_F$ is defined as
\begin{equation}
{\cal E}_F = - r \log r - (1-r) \log (1-r)
\label{supp:e-def-EoF}
\end{equation}
A maximally entangled state has ${\cal E}_F = \log 2$ (and $C = 1$).

\paragraph*{Logarithmic negativity ${\cal E}_N$.} In the eigenbasis of ${\cal H}_0$, take the partial transpose $\rho^\Gamma$ of the density matrix, i.e. $\langle ab | \rho^\Gamma | cd \rangle = \langle ad | \rho | cb \rangle$. (This amounts to swapping $\tau_+$ and $\tau_-$ in the second factor of the tensor products $\tau_\alpha \otimes \tau_\beta$.)
Compute the eigenvalues $\mu_i$ of $\rho^\Gamma$ and sum the negative ones to get the negativity $N = \sum_{\mu_i < 0} |\mu_i|$. Then we have $0 \le N$ and the logarithmic negativity is \begin{equation}
{\cal E}_N = \log(1 + 2 N).
\label{supp:e-def-logNeg}
\end{equation}
A maximally entangled state of two spin-$1/2$ systems also has ${\cal E}_N = \log 2$.

The plots of Fig.\,\ref{dark}\,(right) in the main paper illustrate the phenomenology of quantum correlations at the symmetric point. The states between the triplet mixture $\rho_{\rm tr}$ and the diagonal state $\rho_{\rm st}$ are (nearly everywhere) correlated. Although the bright state $|b\rangle$ is maximally entangled, its weight is not sufficient to guarantee non-classical behaviour, that would become manifest when for example Bell inequalities are violated. Going towards the maximally entangled state, both quantum measures ${\cal E}_F$ and ${\cal E}_N$ signal an entangled state when the weight $p$ of the dark state exceeds $\sim 0.5$. Eventually, both measures converge for $p \to 1$ to the same value (as they must for maximally entangled states).

We conclude this section by calculating the minimum mixing parameter $p_c$ to have an entangled stationary state. The family of density matrices $\rho$ has the same block form~(\ref{supp:eq-effective-H}) as the bath-induced Hamiltonian. $\rho$ and its partial transpose may be written
\begin{equation}
\rho = \begin{pmatrix}
e & & & \\
& a & c &
\\
& c & b &
\\
 & & & g
\end{pmatrix}
\,,\qquad
\rho^\Gamma = \begin{pmatrix}
e & & & c \\
& a & &
\\
& & b &
\\
c & & & g
\end{pmatrix}
\label{eq:}
\end{equation}
with real-valued $c = x/2$.
The eigenvalues of $\rho^\Gamma$ are $a, b$ (which are positive) and $\mu_{1,2} = \tfrac12( e + g ) \mp \tfrac12[(e - g)^2 + 4 c^2]^{1/2}$. A necessary condition for entanglement is $\mu_1 \le 0$, equivalent to $c^2 \ge e g$ and setting a lower bound to the off-diagonal element $c$ as expected. (The same condition is found by requiring a nonzero concurrence $C$.)

To translate this into the mixing parameter $p$, we recall that for the dark state projector $|d\rangle\langle d|$, the off-diagonal element at the position of $c$ is $-\frac{1}{2}$. The triplet state $\rho_{\rm tr}$ is diagonal in the basis $\{ |{\uparrow\uparrow}\rangle, |b\rangle, |{\downarrow\downarrow}\rangle \}$ whose populations are $q^2/Z$, $q/Z$, and $1/Z$ with $Z = 1 + q + q^2$ and the `Boltzmann factor' $q = {\rm e}^{ -\beta^* \nu } = (\eV - \nu) / (\eV + \nu)$. The bright state $|b\rangle\langle b|$ has $+\frac{1}{2}$ as off-diagonal element. The parametrisation $\rho = p \, |d\rangle\langle d| + (1-p) \rho_{\rm tr}$ thus yields the expressions
\begin{equation}
c = \frac12 \Big( {- p} + (1-p) \frac{ q }{ Z } \Big)
\,,\qquad
e = (1-p) \frac{ q^2 }{ Z }
\,,\qquad
g = (1-p) \frac{ 1 }{ Z }
\label{eq:}
\end{equation}
The inequality above for $c$ now yields $\tfrac12 \big| p ( 1 + q/Z ) - q/Z \big| \ge (1 - p) q/Z
$. A brief calculation gives the lower limit
\begin{equation}
p \ge p_c = \frac{ 3 q }{ 1 + 4 q + q^2 } = \frac{ \eV^2 - \nu^2 }{ 2 ( \eV^2 - \nu^2/3) }
\label{eq:def-critical-mixing-pc}
\end{equation}
For a large bias, we get $p_c \to \frac12$.
The limit $\eV \to \nu$ ($p_c \to 0$) is not consistent with the approximations behind our model, since we require ${\rm e}^{ -\beta( \eV - \nu ) } \ll 1$, but this does not exclude values of $p_c$ significantly smaller than $\frac12$.

\subsubsection{Transient entanglement}

\begin{figure}[thp]
\centerline{
\includegraphics[height=50mm]{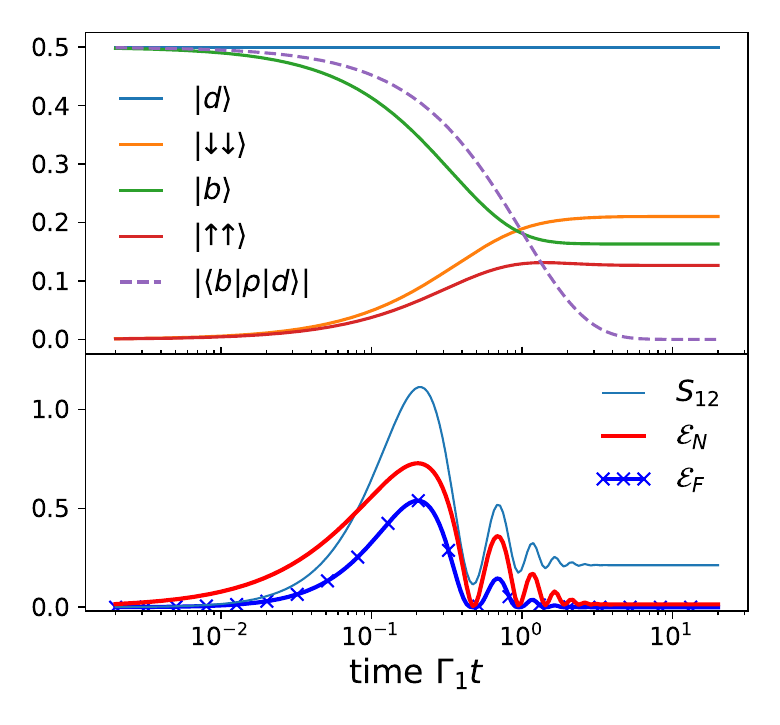}
\hspace*{05mm}
\includegraphics[height=50mm]{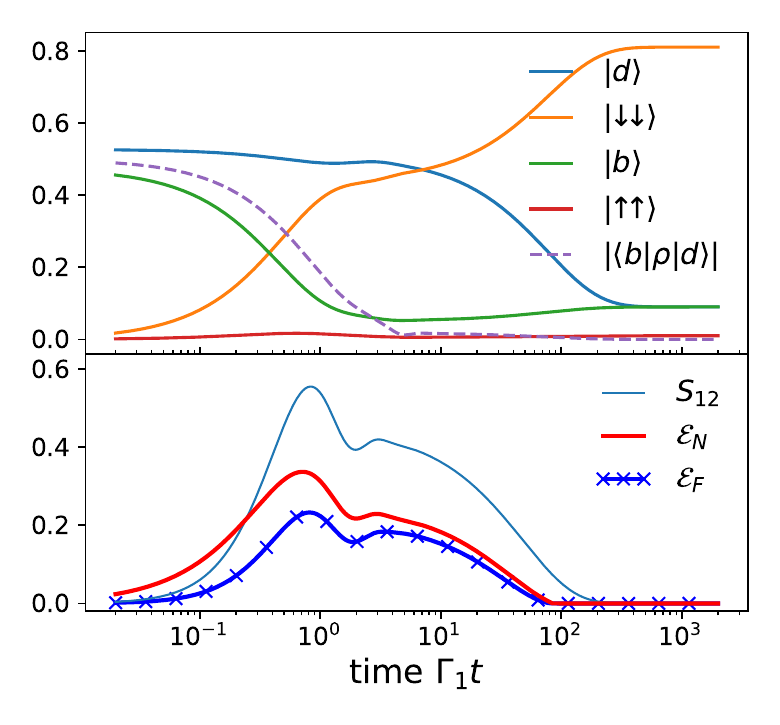}
}
\caption[]{\parbox[t]{140mm}{\raggedright
Transient behaviour of the two-spin system starting from the state $|{\downarrow\uparrow}\rangle$, i.e. one spin excited.
Top panels: populations in the eigenbasis of the effective interaction (the dashed line illustrates the decaying coherence between the dark state and the bright state in the triplet sector).
Bottom panels: entanglement measures and correlation entropy $S_{12}$ of the evolving two-spin density matrix $\rho(t)$.
Entropy and entanglement measures are scaled to (e)bits, using logarithms to base $2$.
(\emph{left}) Symmetric point $\lambda_1 = \lambda_2 = 0.0126$. A relatively strong effective exchange interaction ($J_{\rm ex} \approx 0.164\,\nu \approx 1.64\,\Gamma_1$) leads to oscillations in the entanglement. Its asymptotic value is small because of the high bias voltage $\eV \approx 7.93\,\nu$.
(\emph{right}) Slightly different couplings $\lambda_1 = 1.26 \lambda_2 = 0.0126$
at lower bias $\eV = 1.25\,\nu$ (effectively colder equilibrium temperature), weaker exchange $J_{\rm ex} \approx 0.0054\,\nu \approx 0.34\,\Gamma_1$.
The entanglement vanishes even before the $|d\rangle$ and $|b\rangle$ populations have equilibrated.
In both cases, no spin-orbit coupling.
The time axis is logarithmic and scaled to $\Gamma_1 = 0.1\,\nu$ (left)
and $\approx 0.016\,\nu$ (right).
}}
\label{fig:transient-entanglement}
\end{figure}

We show in Fig.\,\ref{fig:transient-entanglement} two further examples of the transient behaviour of the entanglement in the two-spin system. The initial state ($|{\downarrow\uparrow}\rangle$, one spin excited) is a superposition of the dark and bright state [as in the main paper, Fig.\,\ref{dark}(\emph{right})]. The left panel is taken at the symmetric point (the population of the dark state $|d\rangle$ is constant).
The three triplet states evolve over a time scale $1/\Gamma_1$ towards an equilibrium distribution. Super-imposed on this is an oscillatory behaviour of the entanglement, which we attribute to the relatively strong exchange interaction that operates on the off-diagonal matrix element $\rho_{+-}$.
The asymptotic entanglement is weak (the correlations are mainly classical, as evidenced by $S_{12}$), which is due to the high effective temperature of the triplet system. Indeed, in this setting, the mixing parameter $p$ is constant, while its critical value~(\ref{eq:def-critical-mixing-pc}) is $p_c \approx 0.4947$.

In the right panel, the system is shifted away from the symmetric point. While the triplet levels equilibrate (to a lower temperature), the bright and dark states `decohere', as shown by the matrix element $\langle b|\rho(t)|d\rangle$. The corresponding rate is $\half(\gamma_{+b} + \gamma_{-b} + \gamma_{+d} + \gamma_{-d}) + \gamma_{\rm a} \approx 0.90\,\Gamma_1$, while the exchange interaction, that competes with this decoherence, is chosen weaker here. For $\Gamma_1 t \agt 3$, the system walks roughly through the stationary manifold spanned by the triplet state $\rho_{\rm tr}$ and the dark state. The transient entanglement is larger ($p_c \approx 0.229$ and initially $p \approx 0.527$), but eventually decays as the dark state equilibrates with the triplet. For these parameters, the sum of transition rates leaving $|d\rangle$ is $\gamma_{+d} + \gamma_{-d} + r_{d \to b} \approx 0.015\,\Gamma_1$. The marked vanishing of ${\cal E}_N$ and ${\cal E}_F$ illustrates the ``sudden death of entanglement'' discussed previously for two coupled oscillators coupled to a bath.\cite{sm:Paz_2008}

We have found numerically that over time scales as long as in the Figures, a stable solution of the system dynamics can be obtained using the eigenvalues and eigenvectors of the linear operator that represents the master equation. This technique is also used for correlation functions and spectra, see Sec.\,\ref{supp:D}.

\subsection{Additional interactions and spectra}

\subsubsection{Dipole-dipole interaction}
\label{sm:dd-interaction}

We consider here an additional interaction between the two spins ${\cal H}_{12}$. This Hamiltonian can be treated in two ways:~\cite{sm:gonzalez,sm:shavit} (i) Either dissipative terms are evaluated for the original system coupled to its environment, and then ${\cal H}_{12}$ is added to the Hamiltonian part ${\cal H}'$ of the master equation (``local method''). (ii) Or ${\cal H}_{12}$ is included into the bare system Hamiltonian before evaluating the dissipative terms. This typically requires the diagonalization of ${\cal H}_0 + {\cal H}_{12}$ (``global method''). If the energy scale $D$ measures the strength of ${\cal H}_{12}$,
the global method fails in the limit $D \to 0$ which conflicts with its secular approximation, as additional degeneracies appear.
If, on the contrary, $D$ becomes large, the local method fails when the dissipative terms (bath correlation spectra) are evaluated at incorrect frequencies (in the generic case of a ``structured bath''). The crossover of validity between the two methods is estimated \cite{sm:gonzalez} to occur when $D$ is comparable to the level spacing in ${\cal H}_S$. We note that the previous bath-induced interaction ${\cal H}'$ does not fit into either method, yet this is valid even for moderate $J_{\rm ex}$ or $J_{\rm DM}$ since these particular interactions originate directly from the system-environment coupling. Yet, the condition for the secular approximation is that the linewidths be much smaller than $\nu$, hence from Eqs.\,(\ref{e06}, \ref{e07}) in the main text, $J_{\rm ex}, J_{\rm DM} \ll \nu (\Lambda / \eV)$, a comparable limitation to that of using the local method.

Our example is a dipole-dipole interaction with the simplest symmetry that has rotation invariance around the magnetic field axis.\cite{sm:wertz} It depends on a single parameter $D$ such that the interaction is
\beq{0912}
{\cal H}_{\rm dip} = \frac{D}{3} \big(
    \tau_z\otimes\tau_z - \tau_+\otimes\tau_- - \tau_-\otimes\tau_+
\big)
\eeq
We study this coupling by adding ${\cal H}_{\rm dip}$ to the effective Hamiltonian ${\cal H}'$, i.e. the ``local'' method. This is essential for treating properly the Larmor degeneracy as well as keeping the presence of a dark state. The practical validity of this approach is expected \cite{sm:gonzalez} to be $D/3\lesssim \nu$. The spectrum of ${\cal H}_0 + {\cal H}_{\rm dip}$ has triplet states at $\pm\nu + \frac{1}{3}D,\,-\frac{2}{3}D$ and a singlet state at energy $0$. The allowed ESR transitions are within the triplet states at $|\nu\pm D|$, while an STM experiment would show also the singlet-triplet transitions, i.e.\ additional lines at $|\nu\pm\frac{1}{3}D|$. This phenomenon can be confirmed by the spectra shown in Fig.\,\ref{dipole}.

\begin{figure} \centering
\includegraphics*[width=.7\textwidth]{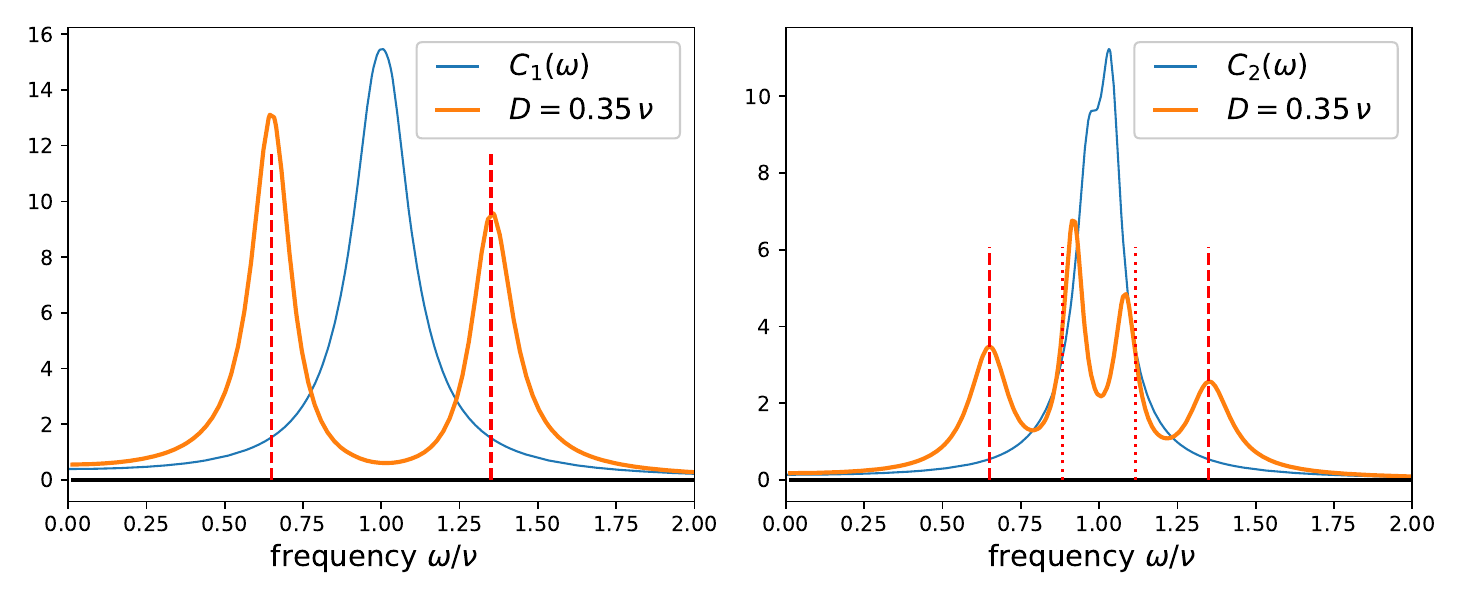}
\caption[]{\parbox[t]{140mm}{\raggedright
Correlations of Eq.\,\eqref{e11} in the main text, including a dipole-dipole interaction [Eq.(\ref{supp:e0912})] with strength $D = 0.35\,\nu$ (thick solid lines). Thin lines: $D = 0$. Left: $C_1(\omega)$, right: $C_2(\omega)$. The vertical lines mark the expected positions as explained in the text. Parameters: coupling constants $\lambda_1 = \lambda_2 \approx 0.005$, $V \approx 7.93\,\nu$. We have chosen exchange and DM interactions, $J_{\rm ex} \approx 0.00854\,\nu$ and $J_{\rm DM} \approx 0.00381\,\nu$, much weaker than the dipole-dipole coupling.
}}
\label{dipole}
\end{figure}

\subsubsection{Regression formula and spectra}
\label{supp:D}

The spectra shown in the main paper and in Fig.\,\ref{dipole} arise from two-time correlation functions $\langle A(t')B(t)\rangle$ that we compute using the quantum regression formula.\cite{sm:martin, sm:shavit} We introduce the super-operator $R$ that represents the master equation $d\rho/dt = R \rho$ when the density matrix is represented as a vector. This provides formally the solution $\rho(t) = \exp( R t ) \rho(0)$ and the stationary state $\rho_{\rm st}$ is the solution of $R\rho_{\rm st}=0$. The regression formula provides the expression ($t' > t$)
\begin{equation}
\langle A(t')B(t)\rangle = \tr\big\{ A \exp[ R (t'-t) ] B \rho_{\rm st} \big\}
\label{supp:e-regression-formula-A(t)B}
\end{equation}
where it is understood that the correlation is evaluated in the stationary state. For $t' < t$, we may use $\langle A(t')B(t) \rangle = \langle B^\dag(t) A^\dag(t') \rangle^* = \langle A(t) B(t') \rangle^*$ for the operator pairs $A = B^\dag$ considered in the main text.
The Fourier transform with respect to $t' - t$ can thus be written as
\begin{equation}
\langle A(t')B(t)\rangle_\omega = -2\re\tr[A\frac{1}{R+i\omega}B\rho_{\rm st}]
\label{supp:e-regression-spectrum}
\end{equation}
This can be evaluated conveniently from the eigenvalues and eigenvectors of the super-operator $R$. An eigenvalue $\mu$ with a vanishing real part gives rise to a $\delta$-peak centered at $-\im\mu$. Its weight can be found from the corresponding eigenvector.

For the representation of the super-operator $R$, a matrix notation is convenient that we illustrate here for a single spin $1/2$ with the basis states $|a\rangle = |0\rangle, |1\rangle$. Using a tilde for the matrix elements $\tilde\rho_{ab},\,a,b = 0,1$ in the eigenbasis, we map them to a four-vector~\cite{sm:shnirman1}
\beq{79}
\left(\begin{array}{cccc} 1 & 0 & 0 & 1\\\half & 0 & 0 & -\half \\ 0 & 1 & 0 & 0 \\
0 & 0 & 1 & 0 \end{array} \right)\left(\begin{array}{c} \tilde\rho_{00}\\ \tilde\rho_{01} \\ \tilde\rho_{10} \\ \tilde\rho_{11} \end{array} \right)=\left(\begin{array}{c} \rho_0\\ \rho_z \\ \rho_+ \\ \rho_- \end{array} \right)
\eeq
with $\rho_0 = 1$.
The system Hamiltonian is ${\cal H}_S=\half\nu\sigma_z$ where $\nu$ is the Larmor frequency and $\sigma_x, \sigma_y, \sigma_z$ are Pauli matrices. They determine the magnetization ${\bf M}=\tr[\bfsigma \tilde\rho]$.

The standard Bloch equation, as derived in Eq.\,\eqref{supp:e2501},
\beq{80}
&& \dot\rho_+=-i\nu\rho_+-\frac{1}{T_2}\rho_+,\qquad  \dot\rho_z=-\frac{1}{T_1}(\rho_z-\rho_z^0)
\eeq
identifies $R$ as
\beq{801}
R=\left(
    \begin{array}{cccc}
0 & 0 & 0 & 0 \\
\rho_z^0/T_1\qquad & -1/T_1 & 0 &  0 \\
0 & 0 & -i\nu -1/T_2 & 0 \\
0 &  0 & 0 & i\nu -1/T_2
\end{array} \right)
\eeq
To evaluate correlation functions we need the Pauli matrices (or any other operator) as $4\times4$ matrices that operate on the 4-vector $\rho$. The action of the $2\times2$ matrix $A$ acting from the left, $A \tilde\rho = B$ gives
\beq{81}
&&\left(\begin{array}{cc} A_{00} & A_{01}\\ A_{10} & A_{11} \end{array} \right)
\left(\begin{array}{cc}\half+\rho_z & \rho_+\\ \rho_- & \half-\rho_z \end{array} \right)=
\left(\begin{array}{cc} A_{00}(\half+\rho_z)+A_{01}\rho_- & \qquad A_{00}\rho_+ +A_{01}(\half-\rho_z)\\
A_{10}(\half+\rho_z)+A_{11}\rho_- & \qquad A_{10}\rho_+ +A_{11}(\half-\rho_z)\end{array} \right)
\eeq
Mapping the elements $B_{ab}$ to a 4-vector, i.e.\ $(\tr B = B_{00}+B_{11}, \half(B_{00}-B_{11}), B_{01},B_{10})$ as in Eq.\,(\ref{supp:e79}), we get
\beq{8101}
&&
\left(\begin{array}{c}  A_{00}(\half+\rho_z)+A_{01}\rho_- +A_{10}\rho_+ +A_{11}(\half-\rho_z) \\
\half[A_{00}(\half+\rho_z)+A_{01}\rho_- -A_{10}\rho_+ -A_{11}(\half-\rho_z)] \\A_{00}\rho_+ +A_{01}(\half-\rho_z)\\
A_{10}(\half+\rho_z)+A_{11}\rho_- \end{array} \right)
\nonumber
\\
&&\qquad\qquad\qquad=
\left(\begin{array}{cccc} \half(A_{00}+A_{11}) & A_{00}-A_{11} & A_{10} & A_{01}\\
\frac{1}{4}(A_{00}-A_{11}) & \half (A_{00}+A_{11}) & -\half A_{10} & \half A_{01} \\
\half A_{01} & -A_{01} & A_{00} & 0 \\
\half A_{10} & A_{10} & 0 & A_{11} \end{array} \right)
\left(\begin{array}{c} 1 \\ \rho_z \\ \rho_+ \\ \rho_- \end{array} \right)  \nonumber
\eeq
The last matrix identifies the $4\times4$ form of the original $2\times2$ matrix $A$. Hence the Pauli matrices when operating from the left become
\beq{82}
\sigma_z=\left(\begin{array}{cccc} 0&2&0&0\\\half &0&0&0\\ 0 &0&1&0\\0&0&0&-1 \end{array} \right)
\qquad
\sigma_+=\left(\begin{array}{cccc} 0&0&0&1\\0&0&0&\half\\ \half &-1&0&0\\0&0&0&0 \end{array} \right)
\qquad \sigma_-=\left(\begin{array}{cccc} 0&0&1&0\\0&0&-\half&0\\ 0 &0&0&0\\\half&1&0&0 \end{array} \right)
\eeq
Following the process above for matrices acting from the right, $\tilde\rho A$, we obtain
\beq{84}
\sigma'_z=\left(\begin{array}{cccc} 0&2&0&0\\\half &0&0&0\\ 0 &0&-1&0\\0&0&0&1 \end{array} \right)
\qquad
\sigma'_+=\left(\begin{array}{cccc} 0&0&0&1\\0&0&0&-\half\\ \half &1&0&0\\0&0&0&0 \end{array} \right)
\qquad \sigma'_-=\left(\begin{array}{cccc} 0&0&1&0\\0&0&\half&0\\ 0 &0&0&0\\\half&-1&0&0 \end{array} \right)
\eeq
To illustrate this technique we consider the transverse spin correlation function and find (using Mathematica)
\beq{78}
&&C_{-+}(\omega)=\int\! dt\,\langle\sigma_-(t)\sigma_+(0)\rangle \eexp{i\omega t}=
\frac{(1-2\rho_z^0)/T_2}{(\omega-\nu)^2+(1/T_2)^2}
\nonumber
\\
&&
A(\omega)=\omega[C_{-+}(\omega)-C_{+-}(-\omega)] =\frac{-\rho_z^0\omega/T_2}{(\omega-\nu)^2+(1/T_2)^2}
\eeq
where the absorption rate $A(\omega)$ is a known result for the Bloch equations
[see e.g., Eq.\,(2.48) of Ref.\,\onlinecite{sm:abragam}].

\end{widetext}

\end{document}